\documentclass[useAMS,usenatbib]{mn2e}

\usepackage{graphicx}
\usepackage{times}
\usepackage{natbib}
\usepackage{color}

\newcommand{\apjs}{ApJS}
\newcommand{\apj}{ApJ}

\newcommand{\mnras}{MNRAS}
\newcommand{\aap}{A \& A}

\newcommand{\nat}{Nature}

\newcommand{\memsai}{Memorie della Societ\'{a} Astronomica Italiana}

\newcommand{\enzo}{\it{\small ENZO}}

\begin{document}

\title[Vorticity in Galaxy Clusters]{Turbulence and Vorticity in Galaxy Clusters Generated by Structure Formation}
\author[F.Vazza, T. W. Jones, M. Br\"{u}ggen , G. Brunetti, C. Gheller, D. Porter, D. Ryu]{F. Vazza$^{1}$\thanks{Email:franco.vazza@hs.uni-hamburg.de}, T. W. Jones$^{2}$, M. Br\"{u}ggen$^{1}$, G. Brunetti$^{3}$, C. Gheller$^{4}$, D. Porter$^{2}$, D. Ryu$^{5,6}$\\
$^{1}$ Hamburger Sternwarte, Gojenbergsweg 112, 20535 Hamburg, Germany\\
$^{2}$ University of Minnesota Twin Cities Minneapolis, MN, USA\\
$^{3}$ Istituto di Radioastronomia, INAF, via Gobetti 101, Bologna, Italy\\
$^{4}$ Swiss National Supercomputing Centre Lugano, Switzerland \\
$^{5}$ Department of Physics, UNIST, Ulsan 44919, Korea Ulsan, Korea \\
$^{6}$ Korea Astronomy and Space Science Institute, Daejeon, 34055, Korea \\
}
\date{Received / Accepted}

\maketitle


\begin{abstract}
Turbulence is a key ingredient for the evolution of the intracluster medium, whose properties can be predicted with high resolution numerical simulations.
We present initial results on the generation of solenoidal and compressive turbulence in the intracluster medium during the formation of a small-size cluster using highly resolved, non-radiative cosmological simulations, with a refined monitoring in time. In this first of a series of papers, we closely look at one simulated cluster whose formation was distinguished by a merger around $z \sim 0.3$. We separate laminar gas motions, turbulence and shocks with dedicated filtering strategies and
distinguish the solenoidal and compressive components of the gas flows using Hodge-Helmholtz decomposition.  Solenoidal turbulence dominates the dissipation of turbulent motions ($\sim 95\%$) in the central cluster volume at all epochs. The dissipation via compressive modes is found to be more important ($\sim 30 \%$ of the total) only at large radii ($\geq 0.5 ~r_{\rm vir}$) and close to merger events.  
We show that enstrophy (vorticity squared) is good proxy of solenoidal turbulence. All terms ruling the evolution of enstrophy (i.e. baroclinic, compressive, stretching and advective terms) are found to be significant, but in amounts that vary with time and location. Two important trends for the growth of enstrophy in our simulation are identified: first,  enstrophy is continuously accreted into the cluster from the outside, and most of that accreted enstrophy is generated near the outer accretion shocks by baroclinic and compressive processes.  Second, in the cluster interior vortex stretching is dominant, although the other terms also contribute substantially.

\end{abstract}




\section{Introduction}
\label{sec:intro}

The rarefied media in galaxy clusters (ICMs) are highly dynamic and likely to be
turbulent, with strong motions on many scales that can significantly influence a wide range of ICM physical processes \citep[e.g.,][]{2006PhPl...13e6501S,su06,bl07,2011MmSAI..82..588J}.
These motions may be driven by processes originating on galactic scales (e.g., star burst winds, AGN outflows and bubbles, \citep[e.g.,][]{2009ApJ...694.1317O,2010MNRAS.407.1277M,2012ApJ...750..166M,2012ApJ...746...94G}), possibly ICM-based magneto-thermal instabilities \citep[e.g.,][]{2011MNRAS.410.2446K,zu13}, but especially by cluster-scale processes associated with cluster formation out of cosmological, large-scale structure 
\citep[e.g.,][]{do05,va06,ry08,lau09,va11turbo,zu11,miniati14,sc14}. 

The resulting ICM driving motions on
scales that range up to at least 100s of kpc will generally include weak-to-moderately-strong shocks and hydrodynamic shear, both of which are
expected to lead to turbulent motions that cascade downwards towards dissipation scales.

The solenoidal motions will stretch and fold structures, so are
primarily responsible for amplifying and tangling the ICM magnetic field \citep[e.g.,][]{pjr15,bm15}. The compressive turbulence component will, itself, produce weak shocks that can, in turn, generate solenoidal motions \citep[e.g.,][]{pjr15}. 
Both compressive and solenoidal turbulent components may accelerate cosmic rays through second-order Fermi processes \citep[e.g.,][]{2003ApJ...584..190F,bb05,bl07,2016MNRAS.458.2584B}.

Several previous simulation efforts have measured the energy ratio between compressive and solenoidal motions in the ICM, finding a predominance of solenoidal motions  \citep[e.g.,][ ]{ry08,iapichino11,va14mhd}. Interplay between the turbulence and shocks may be important in other respects, as well. For instance, turbulent amplification of magnetic fields by shocks and associated second-order Fermi acceleration leading to radio relic emission has been explored in several recent studies \citep[][]{ib12,do16,ji16,fujita15,fujita16, do16}.
The relative contributions from solenoidal and compressive turbulent components will depend on the manner in
which the turbulence is generated \citep{fed10,pjr15} and its intensity \citep{vazq94}. In the ICM, each of these conditions is 
likely to vary significantly in both space and time. 

The present work is motivated particular by the primary need to establish
{\it when, where, how and at what level} the two turbulence components are produced and what is their relation to cluster formation dynamics. Here we focus on the turbulence itself, postponing its applications to subsequent works.
We focus on turbulence generation, both solenoidal and compressive, and its connections to local ICM dynamical conditions. This complements previous simulation studies that
have examined the global energetics of ICM turbulence evolved during
cluster formation, including its association with major merger activity.

Particularly when issues such as magnetic field amplification and ICM dissipative processes, including cosmic-ray
acceleration, are involved and when their dependences on local conditions are important \citep[e.g.][]{su06,bj14}, it can be
essential to separate solenoidal from compressive turbulent motions.
For example, in recent work \citet{miniati15} showed that the cluster-wide ICM compressive turbulence component
is likely to have a steep (Burgers-law-like) spectrum, greatly reducing the power available for cosmic
ray acceleration compared to a Kraichnan-like spectrum unless that power can cascade to very small scales, where it can more efficiently transfer energy to the cosmic rays. 

In order to establish and evaluate the physical roles of turbulence it is essential to separate truly turbulent, uncorrelated flows from correlated, large scale bulk motions and shocks. Uncorrelated flows cascade energy and vorticity to small scales where they work to amplify magnetic fields and dissipate into heat and nonthermal particle energy. Coherent flows, on the other hand, carry signatures of global dynamical events, but are less directly connected to dissipation and magnetic field development. 

Power spectra and structure functions constructed from simulation cluster-wide velocity fields typically suggest outer coherence
scales $\sim 1$ Mpc \citep[e.g.,][]{vbk09,miniati14}. While these scales correctly capture dominant, energy containing
processes for the entire cluster, they do not, as emphasized above, necessarily discriminate against non-random, so, non-turbulent
motions. They also span highly inhomogeneous, often stratified volumes whose motions on moderate to
small scales are often too separated
to be well connected causally when local driving conditions vary abruptly in response to nonspherical accretion or interactions (including mergers) with halos.
So the ability of such global statistics to represent
turbulent motions on the scales where they are most influential is limited. In that context, a more ``local'' approach
seems better motivated. One strategy of this kind
was suggested by 
\citet{va12filter} and \citet{va14mhd}. We will follow this strategy here in order
to understand more clearly the generation, evolution and dissipation of the solenoidal and compressive turbulent motions
produced during cluster formation.
The following section outlines our simulations. Section 3 provides a summary of the several analysis tools we employ in this work, while Section 4 presents results of these analyses applied to a selected cluster simulation. Section 5 provides a brief summary and conclusion.


\section{Simulations}
\label{sec:methods}

We carried out, using the publicly available {\enzo} code \citep[][]{enzo14},  multiple cosmological simulations designed to follow closely the formation of clusters selected to have a broad range of evolutionary histories. 

The simulations applied the WMAP7 $\Lambda$CDM cosmology \citep[][]{2011ApJS..192...18K}, with $\Omega_0 = 1.0$, $\Omega_{\rm B} = 0.0445$, $\Omega_{\rm DM} =
0.2265$, $\Omega_{\Lambda} = 0.728$, Hubble parameter $h = 0.702$, with
a normalization for the primordial density power spectrum of $\sigma_{8} = 0.8$ and a primordial index of $n=0.961$. All runs were non-radiative. No non-gravitational sources of heating were present {\it except} for an imposed temperature floor of $T = 3 \cdot 10^4$ K in the redshift range $4 \leq z \leq 7$, tuned to mimic the effects of reionization at low redshift.

We generated initial conditions (IC) separately at $z = 30$ for each individual simulation
using 2 levels of nested volumes with comoving dimension, $L_0 = 44~{\rm Mpc}/h = 62.7 (\approx 63)$Mpc. This technique is the same as
introduced in \citet{2007ApJ...665..899W}. 

First, low-resolution runs of several independent cosmological volumes were 
investigated in order to select the most massive objects in the volume.
Second, new IC were generated by nesting two grids of 
 $400^3$ cells each and two levels of DM particles ($400^3$ each) with
increasing mass resolution. The total volume was rotated in order to host
the formation of the pre-selected cluster at the centre of the domain.

Further details of the IC so generated are as follows:

\begin{itemize}
\item level ``0'': resolution $=110/h~{\rm kpc} \approx$ 157 kpc and DM mass resolution of $m_{\rm dm} =8.96 \cdot 10^7 \rm M_{\odot}$, covering the full, co-moving 63 Mpc;
\item level ``1'':  resolution $=55/h~{\rm kpc} \approx$ 78.4 kpc and DM mass resolution of $m_{\rm dm}=1.12 \cdot 10^7 \rm M_{\odot}$, covering the innermost $\approx$ 31 Mpc (centred on the cluster formation region).
\end{itemize}

Inside the central $(L_0/10)^3$ volume of each box ($=6.27 \approx 6.3~\rm{Mpc})^3$ (comoving), which is large enough to include the virial radii of most of our clusters, we further refined the grid by a factor 4. That
increased our innermost spatial resolution to $\Delta x = 13.8/h ~{\rm kpc} \approx 20 ~\rm kpc$. 

The generation of shocks and turbulence in simulated flows may be subject to spurious numerical effects, especially when adaptive
mesh refinement is concerned \citep[e.g.,][]{miniati14,2015A&C.....9...49S}. We wanted to avoid spurious effects caused
by an uneven grid structure over time or space in the cluster formation region;
the imposed fixed mesh refinement scheme puts $100$ percent
of the inner sub-volume uniformly sampled at $14/h~{\rm kpc}~\approx 20$ kpc.
The desired behavior was obtained in {\enzo}  by ``flagging'' all cells within the volume of interest
and using the AMR scheme to compute first the intermediate
level of $40$ kpc, and second the final level spanning the same sub-volume at $20$ kpc.  %
That procedure also ensured a conservative reconstruction
of the fluxes across the coarse boundary of the  $6.3$ Mpc-sized, ``zoom'' region with the 
outer, lower-resolution volume, thus minimizing the noise in the refined
reconstruction of infalling matter from the peripheral regions.
This refinement procedure goes beyond previous sets of simulations by our group, where only a fraction (even if quite large) of the cluster volume was refined with tailored AMR \citep[][]{vbk09}.

The resulting full {\it "Itasca simulated cluster"} (ISC)\footnote{The ``Itasca'' label refers to the HPC cluster at the University of Minnesota used to compute the simulations. The website of the project is accessible at http://cosmosimfrazza.myfreesites.net/isc-project. } sample consists of 20 clusters with the above time and spatial information  and has been designed to let us 
 examine a variety of formation scenarios in detail. Each cluster run required about $\sim 12 000-13 000$ cpu hours (about 1300 root-grid time steps and $\sim 10^5$ top-level time steps in total). For analysis purposes we saved one data cube of hydrodynamical and N-body properties following sequences of 10 root grid time steps before $z=1$, and then after every single root grid time step
for the remaining $z \leq 1$ evolution. This typically lead to $\sim 200$ data dumps being retained for analysis. The dump time resolution after $z = 1$, while not constant, was  generally $\sim 50$  Myr.

In this paper  we limit our analysis to one small-mass cluster, designated as ``cluster it903'', that underwent a major merger event ending around $z \sim 0.3$. At the end of the simulation ($z = 0$), it903 had a total mass, $M_{\rm tot} \approx 10^{14} {\rm M}_{\sun}$, with core and virial radii, $r_{\rm c} \sim 100$ kpc, $r_{\rm vir} \sim 1$ Mpc, respectively. The core temperature, $T_{\rm c} \sim 2\times 10^7$ K, corresponding to a sound speed, $c_{\rm s} \approx 660$ km/sec. 
We defer the analysis of the full ISC cluster sample to future work. 

\subsection{Visual impression of cluster it903}

Figure \ref{fig:fig0} shows the evolution of the total enclosed gas mass and of the mean gas temperature for our "full'' high resolution region ($6.3$ Mpc)$^3$ for cluster it903 and for an inner $(1.44 ~\rm Mpc)^3$, ``cluster-centred'' region moving with that cluster. We use the inner, cluster-centred volume extensively in the following sections for the study of turbulence and enstrophy since it reveals events involving the cluster more distinctly than the larger volume. 
The plot shows the relatively steady mass growth of it903 over cosmological times and the major merger event beginning after $z \approx 0.4$ (followed by a decrease in the central gas mass, due to the outflow of the gas mass initially attached to a merger-involved subunit). This mass history is accompanied by a sharp peak in average gas temperature at $z \approx 0.35$ in the cluster-centred,  zoomed  box and less dramatic, somewhat later ($z \approx 0.3$) temperature spike in the full high resolution region {\footnote{To avoid confusion later on we mention here that in some analyses we employ  a larger, $5.76 ~\rm Mpc^3$, cluster-centred box, since that size roughly matches the virial radius of the final cluster.  For specificity we will refer to this as the ``cluster virial volume''.}}.

Three-dimensional volume rendering snapshots{\footnote{These renderings assign color and opacity values to each voxel in a volume depending on the value of a rendered quantity (e.g., density), then construct perspective views using volume ray casting (http://www.lcse.umn.edu/hvr/hvr.html).} of the $(6.3~{\rm Mpc})^3$ high resolution volume are shown in Fig.~\ref{fig:vrender} at redshifts $z=1$, $z = 0.5$ and $z = 0.32$, which outline the evolution of the cluster into the most significant merger events mentioned above.
The left column of images 
highlights regions of high gas density, so provides a general sense of the mass merger history. 
The centre image column highlights the 3D  shock distribution in this volume at the same times. The shocks are color coded by Mach number for $1.5 \la \mathcal{M} \la 20$ (red-yellow-white). 
The right column then displays the 3D distribution of enstrophy ($\epsilon = $(1/2) vorticity$^2$) as an easy-to-compute and very useful proxy for the turbulence velocity distribution. It is clear from these images, as we discuss in detail below, that the enstrophy distribution is well-connected to the shock distribution, albeit very different in detail.

 \begin{figure}
\includegraphics[width=0.495\textwidth,height=0.8\textwidth]{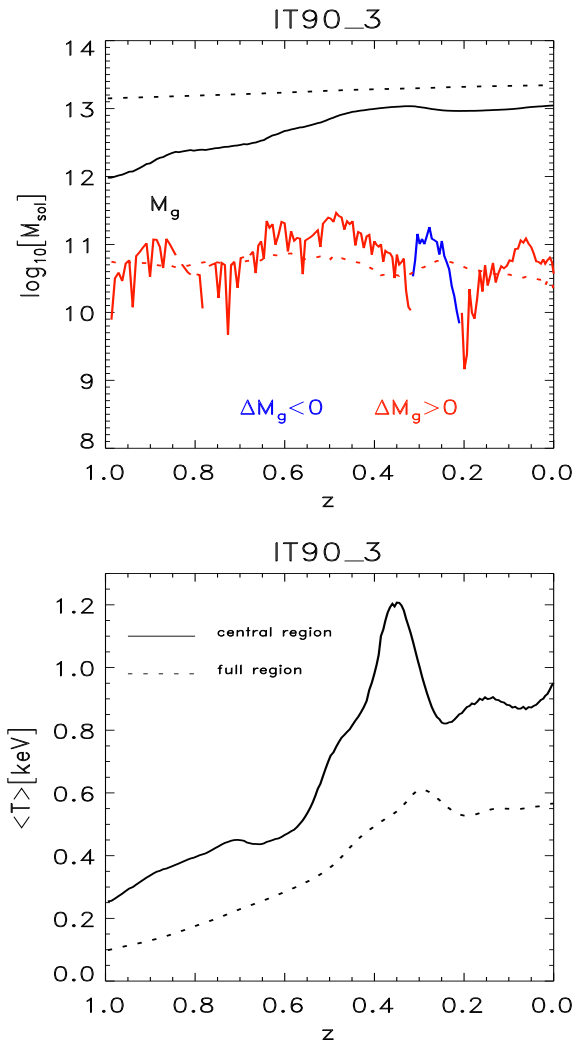}
\caption{Top: evolution of the integrated gas mass (black) and of the gas mass increment/decrement (red/blue) for cluster it903 from z=1 to z=0.  Bottom: evolution of the volume-weighted mean temperature for it903. The solid lines give the evolution for the``innermost'', cluster-centred $(1.44 ~\rm Mpc)^3$ volume used for our primary turbulence analysis, while the dotted lines refer to the full $(6.3 ~\rm Mpc)^3$ peak resolution ($\Delta x \approx 20$ kpc) volume.}
\label{fig:fig0}
\end{figure}
 
 \begin{figure*}
\includegraphics[width=0.95\textwidth]{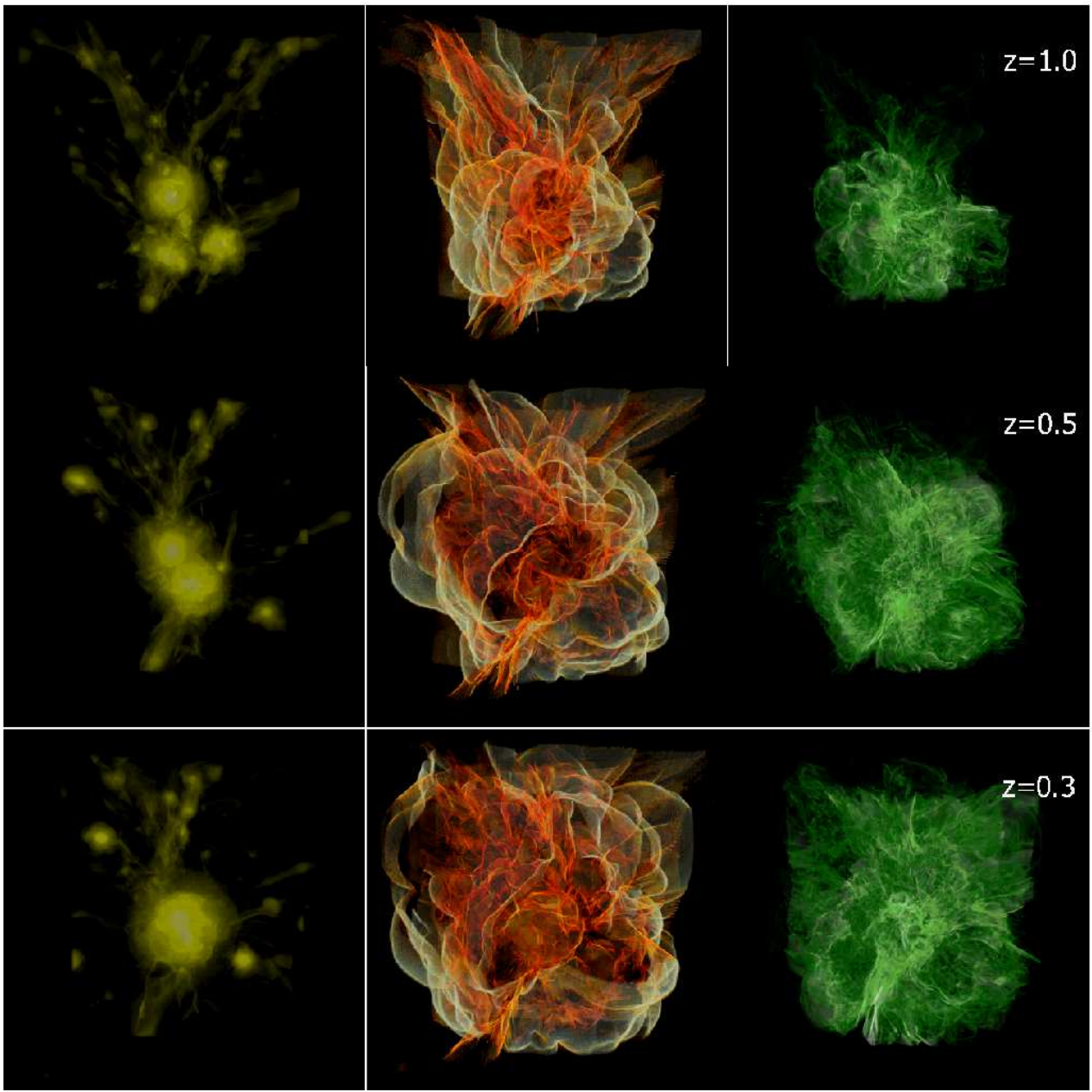}
\caption{Volume renderings of cluster it903 at redshifts z = 1, z = 0.5 and z = 0.32 inside the 6.3 Mpc$^3$ maximum resolution volume with $\Delta x \approx 20$  kpc. Left panels render gas density (arbitrary units). Middle panels show the distribution of shocks color coded by Mach number in the range  ($1.5 \la \mathcal{M} \la 20$ red-yellow-white). Right panels render the enstrophy distribution (arbitrary  units).}
\label{fig:vrender}
\end{figure*}

\begin{figure*}
\includegraphics[width=0.32\textwidth]{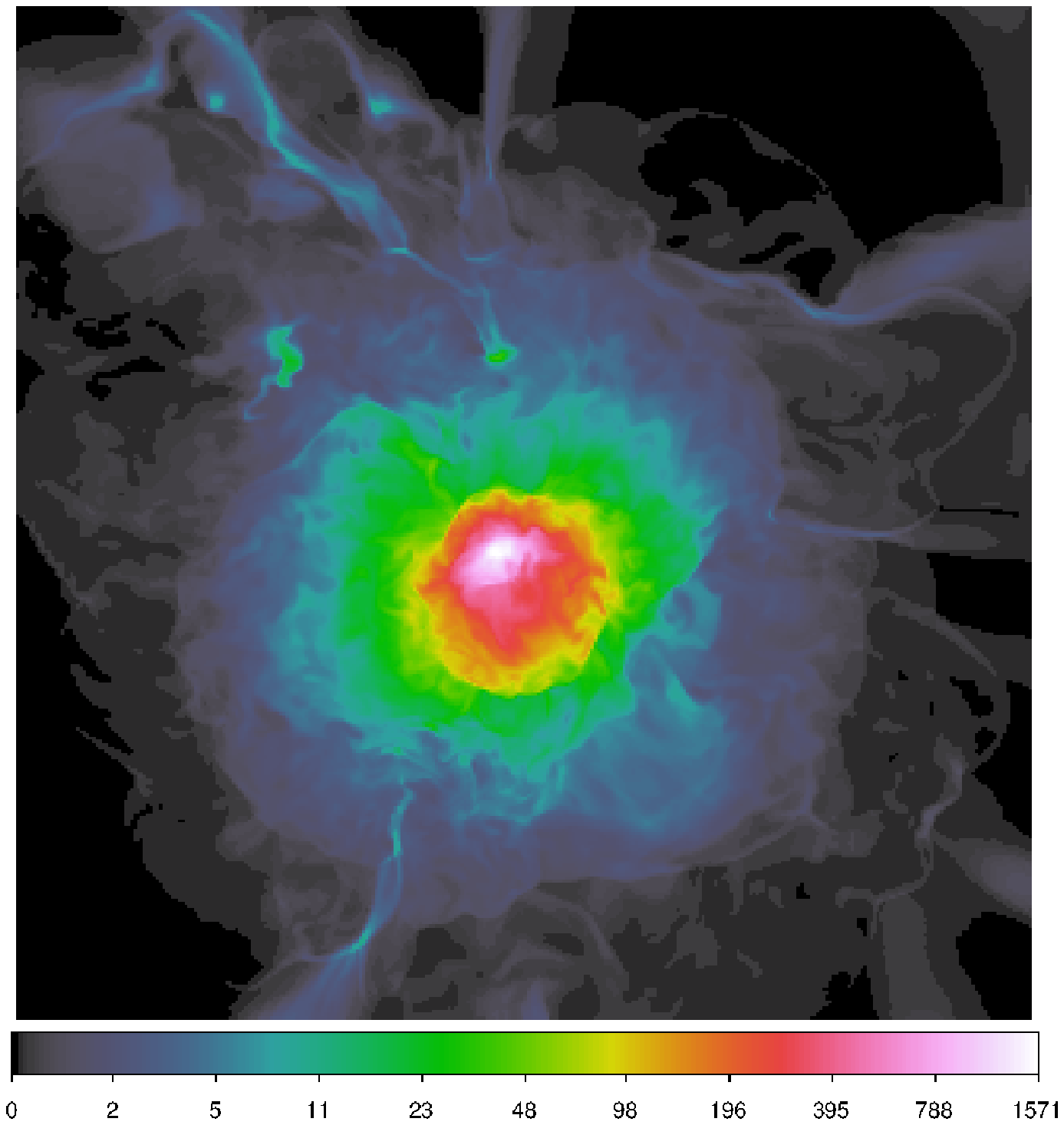}
\includegraphics[width=0.32\textwidth]{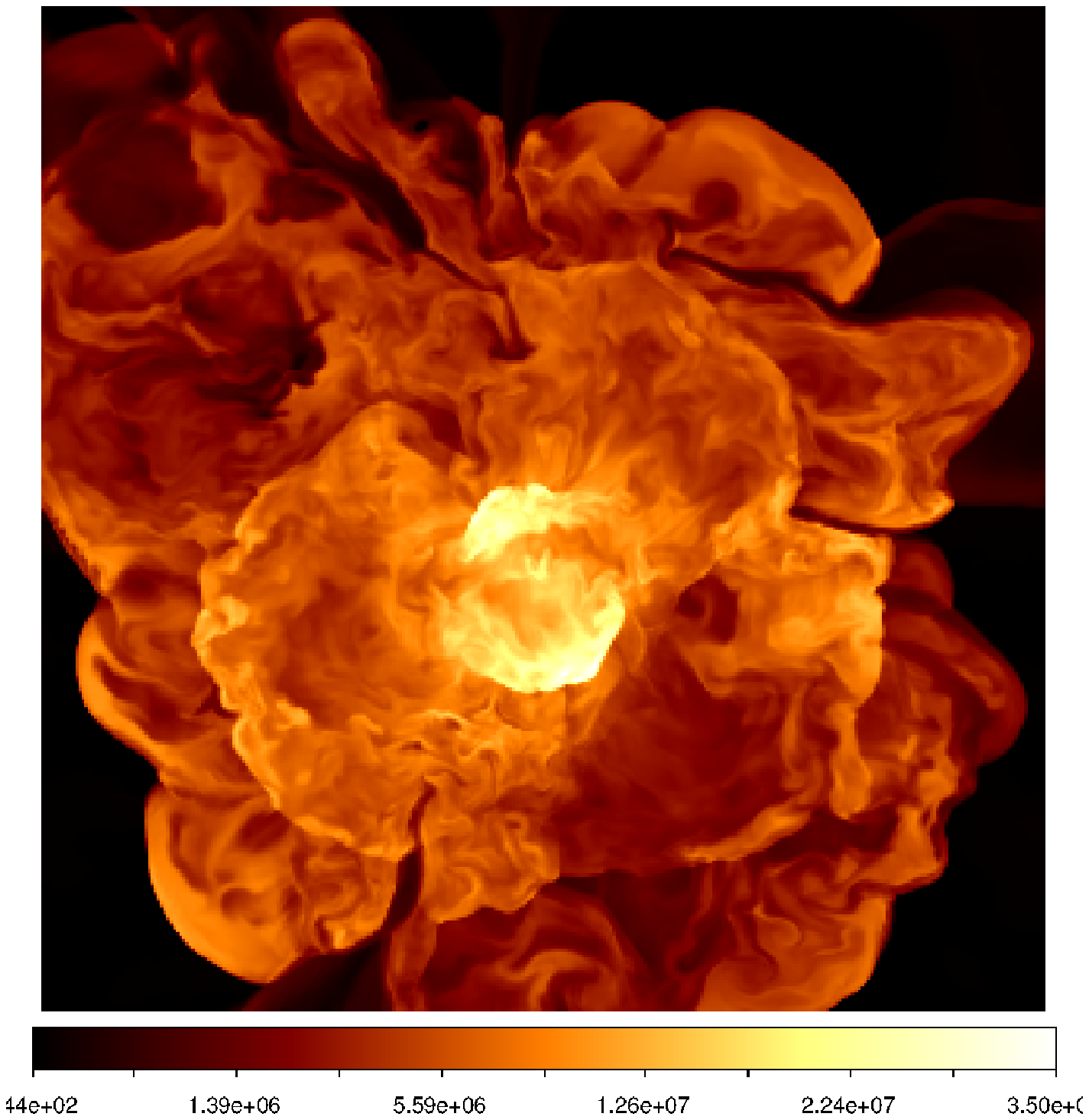}
\includegraphics[width=0.32\textwidth]{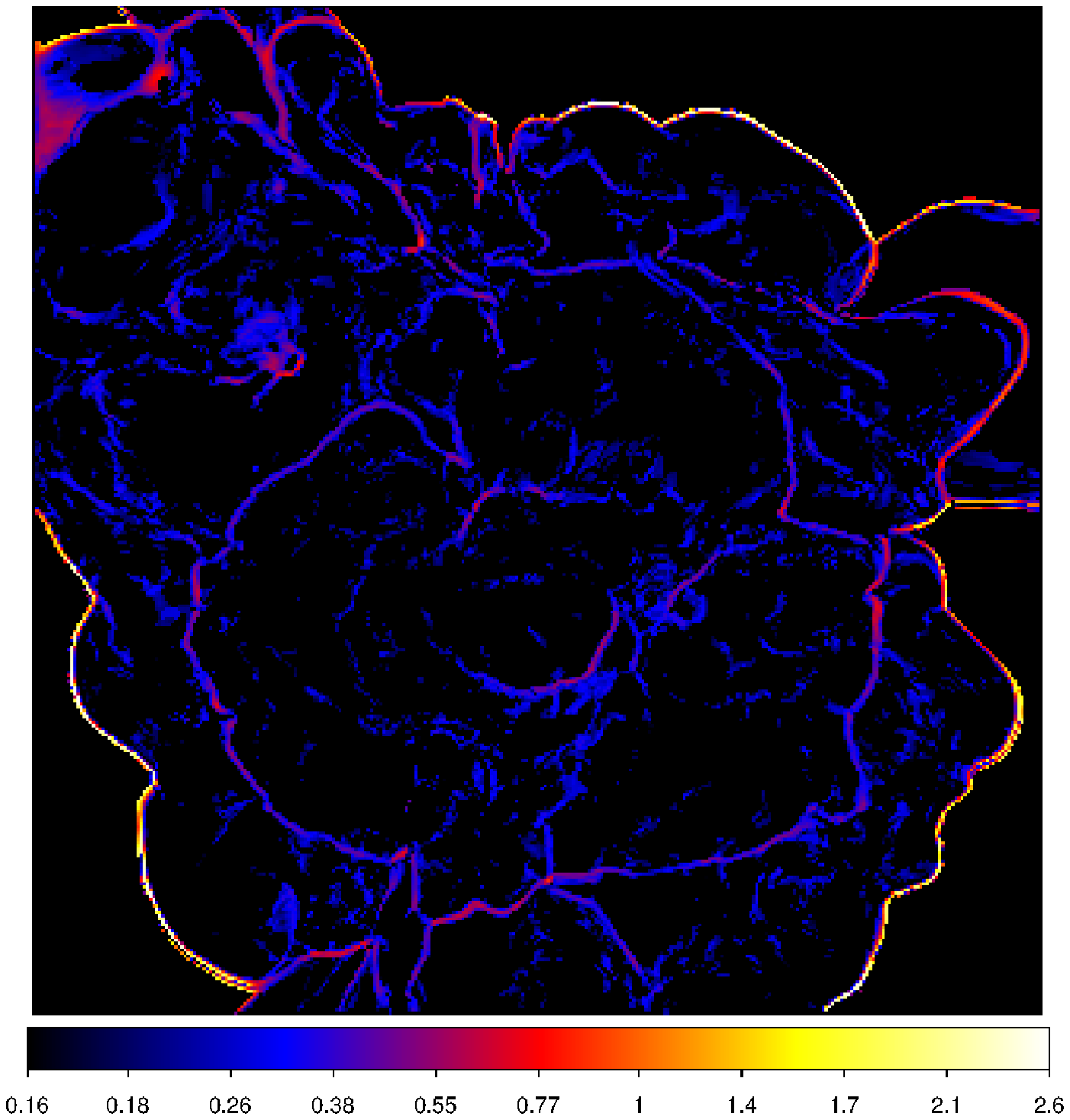}
\caption{Two dimensional slices through the centre of the volume shown in Fig.\ref{fig:vrender} at $z = 0.32$, showing: (left) gas density - units of $[\rho/\langle \rho_{\rm b} \rangle]$, where $\langle \rho_{\rm b} \rangle$ is the mean baryon density)-; (centre) gas temperature -$[\rm K]$-, and the (right) reconstructed map of Mach number -units of $[log_{\rm 10}(\mathcal{M})]$- using our shock finder (Sec.\ref{subsec:shocks}).}
\label{fig:cslice}
\end{figure*}

Fig. \ref{fig:cslice} presents 2D slices at $z = 0.32$ through the centre of the same volume and along the same line of
sight as Fig. \ref{fig:vrender}. The left image shows the gas density, the middle image the gas temperature,
and the right image the shock distribution, again, color coded by Mach number.


\section{Analysis Methods}
\subsection{Shock finder}
\label{subsec:shocks}

We detect shock waves in post-processing analysis using the algorithm presented in 
\citet{va09shocks}. The scheme is based on an analysis of 1D velocity jumps across cells. The minimum of the 3D divergence of the velocity, $\nabla \cdot \vec{v}$, is used to identify the centre of the shock region (see also \citealt{ry03,sk08}). Typically, shock transitions span about 2-3 cells along the shock normal. The one-dimensional Mach number for flagged transition is constructed from the 1D velocity jumps along each scan axis using the Rankine-Hugoniot relations. The final Mach number is constructed from a combination of the three 1D solutions. Shock surfaces are then approximated as the ensemble of the face areas of cells tagged as shocked by the scheme. This method has been extensively tested against similar methods used in grid codes \citep[e.g.,][]{va11comparison}, and has proven to be an efficient and accurate measure of shock waves in cosmological runs.  The kinetic power across each flagged cell shock surface, which provides a useful metric
for energy available to dissipation in shocks, is given by
\begin{equation}
\label{eq:fcr}
f_{\rm KE,shock} = \frac{\rho_{\rm u} v_{\rm sh}^{3}}{2} (\Delta x)^2,
\end{equation}
where $\rho_{\rm u}$ is the co-moving up-stream density, $v_{\rm sh}=\mathcal{M} c_{\rm s}$ is the co-moving speed of the shock, $\mathcal{M}$ is the inflowing Mach number and $\Delta x$ is the cell size. Total kinetic energy flux through shocks is then a sum from Eq. \ref{eq:fcr} over flagged shock cells.

  However,  we point out that the identification and characterization of shocks, and especially shocks with $\mathcal {M} \la 1.5$ and oblique on the grid is made uncertain by the relatively larger numerical errors associated with very small jumps in velocity, by numerical smearing of the shock transition \citep[][]{sk08} as well as by the presence of significant temperature and velocity fluctuations in the ICM, which add uncertainties to estimates of pre-shock values \citep[][]{va09shocks}. In order to bracket the role of, mostly inconsequential, weak shocks in the 
  following turbulent analysis, we will there, also, present complementary results obtained by masking out regions close to identified shocks with $\mathcal{M_{\rm thr}} \ge1.2$. We apply this lower $\mathcal{M_{\rm thr}}$ bound to simplify procedures.  


\subsection{Filtering of turbulent motions}
\label{subsec:turbo}

The extraction of turbulent motions within the cluster 3D velocity field requires  a proper filtering of often comparable coherent velocity components (characteristically larger-scale) from uncorrelated, turbulent velocity components that cascade to small scales.   As noted above, the roles of  solenoidal (rotational, incompressive)  and compressive (irrotational) turbulent motions (defined respectively by $\nabla\cdot\vec { v}_{\rm sol} = 0$ and $\nabla\times\vec {v}_{\rm comp} = 0$)  each have important roles in the ICM. So, as an additional step we also separated the velocity field (both filtered and unfiltered) into those elements.  To accomplish these objectives we combined several steps that we previously proposed and tested individually in \citet{va12filter} and \citet{va14mhd}.

As an initial step to reduce numerical noise and finite difference cross-talk between divergence and curl operations, we applied a first order velocity smoothing filter to the initial full velocity field of the simulation. This has a benign influence on extracted turbulent velocity fields{\footnote{See \citet{pjr15} for a detailed justification of such Favre smoothing operations.}}.
 
For our primary turbulence filter we applied the iterative,
multi-scale velocity filtering techniques based on \citet{va12filter} to the $(6.3~\rm{Mpc})^3$  subvolume of each {\enzo} snapshot.
This filter reconstructs the local mean velocity field around each position, $\vec r$,  by iteratively computing the 
local mean velocity field in the ``n$^{th}$'' iteration as

\begin{equation} 
	\langle{\vec{v}(L_n)}\rangle =\frac{\sum_{i}w_{\rm i}\vec{v_i}} {\sum_{i} w_i},
	\label{eq:turbo}
\end{equation} 
where the sum is over cells within a domain of radius, $ L_n$, and where $w_{\rm i}$ is a weighting function. In this work we simply set $w_{\rm i}=1$ and use a volume-weighting, while in other applications in more stratified media $w_{\rm i}=\rho$ (i.e. density-weighting) is a more appropriated choice. However, given the rather small filtering scales reconstructed by our algorithm in the innermost cluster volume considered in this work (Sec.\ref{sec:results}), the differences between the $w_{\rm i}=1$ and the $w_{\rm i}=\rho$ are very small, as discussed below.

The local small-scale, fluctuating velocity field within the radius, $L_n(\vec r)$, relative to position $\vec r$, is then computed as
$\vec{\delta v}(L_{\rm n}(\vec r))=\vec{v}-\langle{\vec{v}(L_n)}\rangle$ for increasing values of $L_n$.  Iterations are continued until the change in $\delta\vec v$  between two iterations in $L_n$ falls below a given tolerance parameter, which, based on 
our tests, we set to 10\%. 
The resulting $|\vec{\delta v}(L_n)|$ provides our best estimate for the turbulent velocity magnitude
for an eddy-size $L_{\rm eddy} \approx 2 \cdot L_n$.

 We observe that, while in \citet[][]{va12filter} we used the local skewness of the velocity field as a fast proxy to tag shocks, in the present work we can access this information in a more accurate way through the (obviously more computationally intensive) shock finding procedure outlined above (Sec.\ref{subsec:shocks}).  
Therefore, we excluded shocks by simply stopping the iterations whenever a shocked cell entered the domain. That is, the length, $L_n$ then represents the distance to the nearest ``influential'' shock.
On the other hand, our procedure is not designed to explicitly filter out the contribution from velocity shears, e.g. at the contact discontinuity generated by sloshing cold fronts \citep[e.g.][]{2016JPlPh..82c5301Z}. While in principle the presence of such discontinuities might introduce a small spurious contribution to our measured turbulent budget, this spurious signal is small compared to the turbulence induced by mergers \citep[e.g.][]{va12filter}. In particular, the cluster studied in this first paper is a highly perturbed one, where the formation of sloshing cold fronts is highly unlikely \citep[e.g.][]{2016JPlPh..82c5301Z}.

 Our results here, as well as previous cluster simulations are roughly consistent with the behaviour of solenoidal turbulence following the classic, Kolmogorov picture in which $|\delta \vec v| \propto L^{1/3}_n$ (\citep[e.g.,][]{ry08,xu11,va11turbo,miniati14}. Consequently,  while the r.m.s turbulent velocity or the total turbulent pressure depend on $L_n$, the solenoidal kinetic energy cascade flux, defined as:
\begin{equation}
f_{\rm KE,turb}(\vec r) = \frac{F_{\rm KE,turb}}{{\Delta x^3}} = \frac{1}{2}\frac{\rho \delta v(L_n) ^{3}}{L_n},
\label{eq:keflux}
\end{equation}
 is insensitive to the specific value of $L_n$ as long as it is measured within the inertial range  of the turbulence.  
 
Obviously, departures from this behavior can appear if the turbulent behaviour is very different from Kolmogorov and other, distinctive flow patterns are important (e.g., coherent shock waves, see Sec.\ref{subsubsec:shocks}). In the ideal case, our procedure constrains $f_{\rm KE,turb}$ at the outer scale of turbulence, and $L_{\rm turb} = 2 L$, as shown  in the Appendix (Sec. A1.1). However, in practice the iterations are stopped before reaching this scale, and $f_{\rm KE,turb}$  is computed within the turbulent cascade.  For this reason, in the following we will regard $L$ as a {\it filtering} rather than a {\it turbulent} scale.  Reconstruction of the  second is difficult for multiple reasons, including nonequilibrium and highly inhomogeneous flows on large scales. Therefore, in general $L_{\rm turb} \geq 2 L$. As a comparison to the iterative filter we also present below some results in which a simple, fixed scale ($L_f \sim 1$ Mpc) was used.
 
As an additional test, we have verified that the usage of a density-weighting in Eq.\ref{eq:turbo} leaves our results basically unchanged. In particular, the kinetic energy flux measured by Eq.\ref{eq:keflux} is increased only by a factor $\sim 2$ at most, when using the density-weighting within the central $(1.44~ {\rm Mpc})^3$ volume studied in the following (Sec.\ref{sec:results}).


\subsection{Solenoidal \& Compressive Motion Decomposition}
\label{subsec:helm}

In order to characterize the dynamical properties of cluster turbulence we decomposed, both,  the filtered and un-filtered 3D velocity fields of the simulations into solenoidal and compressive elements
using the Hodge-Helmholtz projection in Fourier space \citep[e.g.,][]{kri11}. Here we outline our fiducial method to carry out the decomposition, while in the Appendix (Sec. A1.2) we compare alternative approaches and control tests on our procedure. 

Our fiducial decomposition algorithm first constructed the Fourier space velocity vector field, ${\vec V}(\vec k) = \mathcal{F}(\vec{v}(\vec r))$ using 3D FFTs\footnote{Although this formally assumes the velocity field is periodic in the domain of interest, our tests with non-periodic fields found this to be a minor issue.}, 
then found the solenoidal ($\vec k\cdot{\vec V}_{\rm sol}(\vec k)=0$) component ${V}_{\rm i,sol}, (\vec k), i\in \lbrace 1,3\rbrace$ as,
\begin{equation}
\tilde{V}_{\rm i,sol}(\vec{k}) =
\sum_{j=1}^3 \left( \delta_{i,j} - \frac{k_i k_j}{k^2} \right)
\tilde{V}_j(\vec{k}).
\end{equation}
The compressive component in Fourier space, $\vec k\times\tilde{\vec V}_{\rm comp}(\vec k)=0$,  was found as the residual, $\tilde{V}_{\rm i,comp}(\vec{k})=\tilde{V}_i(\vec{k})-\tilde{V}_{\rm i,sol}(\vec{k})$. Inverse FFTs, $\mathcal{F}^{-1}(\vec V)$, then produced the associated physical solenoidal and compressive velocity distributions, $\vec v_{\rm sol}(\vec r)$ and $\vec v_{\rm comp}(\vec r)$, where, again, $\vec r$ represents a point in the spatial domain. This procedure was performed both on the full, primitive 3D velocity data, yielding  $\vec v_{\rm sol}(\vec r)$ and $\vec v_{\rm comp}(\vec r)$  and on the small-scale filtered field, yielding $\vec{\delta v}_{\rm sol}(\vec r)$ and $\vec{\delta v}_{\rm comp}(\vec r)$. One of our tasks in the present analysis is to compare properties of the two solution sets.

It is useful to point out here that the products of this analysis offer a useful way to estimate the local turbulent energy flux across scales. Specifically, for uncorrelated velocities, $\delta v_L$, filtered on scale, $L(\vec r)$, the turbulent energy flux per unit volume was
estimated using Eq.~\ref{eq:keflux}, for either the compressive,  $\delta v_{\rm L,comp}$, or solenoidal component, $\delta v_{\rm L,sol}$. 
In steady, Kolmogorov turbulence ($\delta v_{L} \propto L^{1/3}$) these fluxes would be scale-independent, so,  would provide robust estimates 
for the local turbulent energy dissipation rate per unit volume, $\rho~\epsilon_d$. 
In the section \ref{subsec:vorticity} we outline an alternate, complementary approach to
estimation of the solenoidal kinetic energy dissipation not requiring the above turbulence scale filtering.


\subsubsection{The influence of shocks in Turbulent Energy Flux Budgets}
\label{subsubsec:shocks}

The presence of cluster formation shocks is problematic to the turbulence component analysis. First of all, the numerically smoothed profiles of shocks contaminate to some degree the solenoidal Fourier field, $\vec{V}_{\rm i,sol}(\vec k)$ for larger $\vec k$ and thus $v_{sol}$ and $\delta v_{\rm sol}$ on small scales. Fortunately, this issue is significantly mitigated  by the velocity smoothing mentioned above, as discussed in \cite{pjr15}, for example. More importantly, while some shocks contribute appropriately to the compressive velocity element, $v_{\rm comp}$, and sometimes to $\delta v_{comp}$, not all shocks, and in particular, structure formation shocks driven by coherent flows, are not elements of uncorrelated, compressive turbulence, $\delta v_{\rm comp}$. The difficult issue, then, is one of judging which shock compressions to count as part of the compressive turbulent motions, $\delta v_{\rm comp}$. In practice, some weak shocks are integral to the turbulence, including those generated by colliding turbulent motions (even solenoidal motions) \citep[e.g.,][]{pjr15}, while other shocks, especially stronger ones and those whose extents exceed the cluster core scales, are more properly associated with the {\it generation} of (random) turbulence \citep[e.g.,][]{fed10,pjr15}, but {\it are not elements} of the turbulence per se. 
We are not aware of any simple, clean and robust way to establish this dichotomy. 
To explore the significance of this complication we carried out a series of numerical experiments in which we masked out patches of cells around shocked cells flagged using methods outlined in Sec. \ref{subsec:shocks}. That specific algorithm is identified below as shock limiting, since, as mentioned above, the length $L_n$ is then limited by the separation scale of shocks with $\mathcal{M}_{thr}$. The resulting kinetic energy statistics were then compared to those obtained without ``shock limiting''. In each case we also examined the velocity fields as extracted from our interative turbulence filter (section \ref{subsec:turbo}) and for turbulence motions obtained using a fixed-scale filter.  
To be conservative with respect to the numerical smearing of shocks, when we applied the masking procedure we removed shock centres, as well as the adjacent $\pm 2$ cells along the shock normal to ensure that the numerical shock profile ($\leq 3$ cells) is fully contained by the masking region. 

\begin{figure}
\includegraphics[width=0.45\textwidth,height=0.4\textwidth]{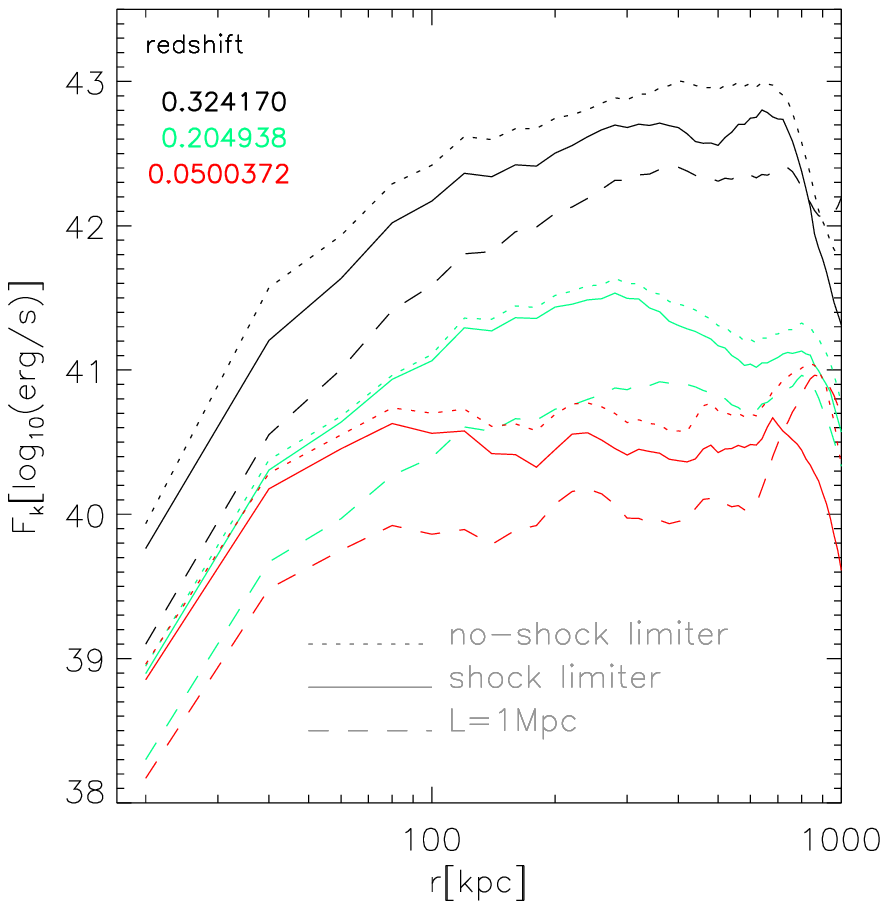}
\includegraphics[width=0.45\textwidth,height=0.4\textwidth]{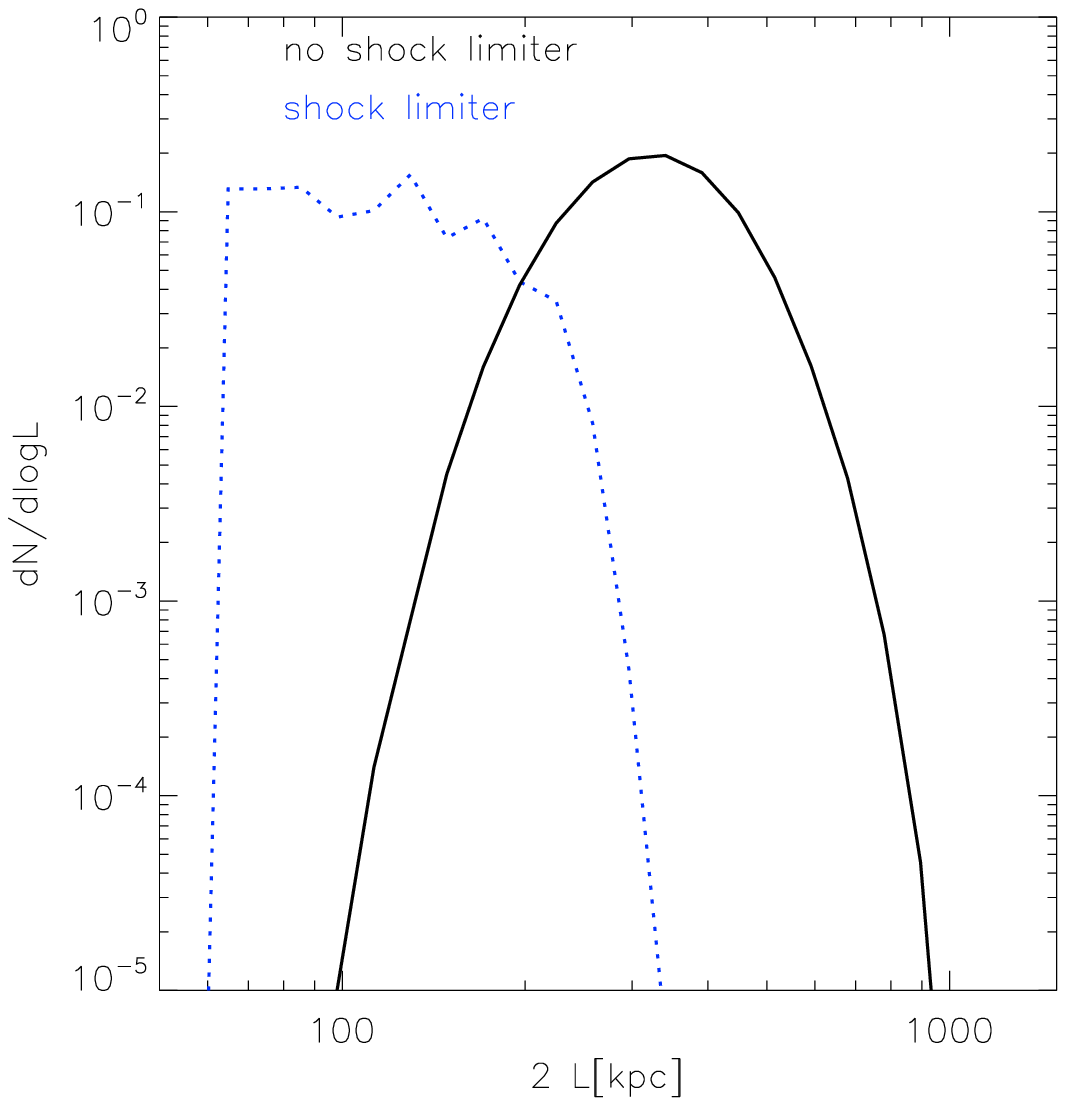}
\caption{Top: radial profiles of turbulent kinetic energy flux (Eq.~\ref{eq:keflux}) for cluster it903 at three epochs, obtained with the iterative filter in (Sec.~\ref{subsec:turbo}). Solid lines  show results with shocks masked out, while for dotted lines no masking was used. The dashed lines show the results obtained using a fixed $L=1$ Mpc filtering scale without shock masking. Bottom: fractional number of cells with a given iterated filtering scales, $2~L$, for the it903 cluster with (dotted curve) or without (solid curve) masking of shocks. The contribution of each cell has been weighted by its turbulent kinetic energy.} 
 \label{fig:fkin_scales}
\end{figure}

While this procedure obviously excludes some kinetic energy on scales of a few cells as well as larger scale flows, we found 
that the net turbulent, kinetic energy fluxes were not sharply reduced by the shock limiting algorithm. As illustrated  in the bottom panel of Fig. \ref{fig:fkin_scales}, the dominant influence of the shock limiter appears to be the reduction in the filtering length, $L_n$. 
Specifically, the shock-limited distribution for $L_n$ is offset to smaller $L_n$ by roughly a factor 3 from the non-shock-limited $L_n$ distribution. On the other hand, the top panel, which presents radial profiles of the kinetic energy flux, $F_{\rm KE}$ (Eq.~\ref{eq:keflux}) demonstrates that the difference between energy fluxes with or without shock limiting is generally less than $\sim~20-40$\%. The application of a cluster-scaled, fixed length turbulence filter, $L_f = 1$ Mpc (comparable to the $r_vir$ of the system), however, led to seriously reduced energy fluxes, typically by factors $\sim~5-10$.
The fact that the fixed filtering scale $L_f=1$ Mpc systematically underestimated the kinetic energy
flux reconstructed in the other approaches suggests that this scale is already larger than the true outer injection scale of turbulence in the domain. This follows from the Kolmogorov picture of turbulent cascades in the ICM, because on scales larger than the injection scale the kinetic energy flux (Eq. ~\ref{eq:keflux}) is not conserved. 
Therefore, the values of r.m.s. velocities measured on these large scales are not truly representative of ICM turbulence. 


\subsection{Enstrophy as a Metric for Solenoidal Turbulence}
\label{subsec:vorticity}

Previous studies \citep[][]{ry08,va14mhd,miniati14,sc14} and our results below suggest that ICM
turbulent motions are predominantly solenoidal in character. 
The distinguishing property of solenoidal turbulence is, of course, that the motions are rotational; that is,
they have non-vanishing local vorticity, $\vec\omega = \nabla\times\vec{v} \ne 0$. The vector vorticity, $\vec\omega$ tends to average towards zero, so the vorticity magnitude is more a more useful tool.
It is useful in this regard to recall that the eddy turn-over rate on a scale $\ell$, 
$1/\tau_{\rm eddy} \sim\delta v_{\rm \ell,sol}/\ell \sim |\omega_{\rm \ell}|$, where the subscript on $|\omega|$ 
identifies this as representing circulation on the specific scale, $\ell$. That is, vorticity is a measure of the rate at which eddies turn over.
We will use this concept below to normalize vorticity measures in convenient units;
i.e., $\hat\omega = \omega \cdot \tau_0$, where $\tau_0$ is a representative timescale.
Then, $\hat\omega$ represents a characteristic number of eddy turn-overs in the chosen  interval, $\tau_0$. 

As an additional perspective, we note that the square of the vorticity, or more directly the enstrophy, $\epsilon = (1/2)\omega^2$, can be related in a turbulent flow to the  kinetic energy content per unit mass, $(1/2) \langle v_{sol}^2\rangle$, and dissipation of the solenoidal turbulence. This measure can then be matched to the solenoidal turbulent energy extracted through the filtering algorithm discussed in Sections \ref{subsec:turbo} and \ref{subsec:helm}. But, since no such filtering is involved in finding the enstrophy, the methods are complementary.

Formally, in terms of the turbulence one-dimensional 
velocity power spectrum, $E_s(k)$,  the mass weighted solenoidal kinetic energy can also be written as:
\begin{equation}
\epsilon = \frac{1}{2}\langle v_{\rm sol}^2\rangle \frac{\int k^2 E_s(k) dk}{\int E_s(k) dk} = \frac{1}{2}\langle v_{\rm sol}^2\rangle \bar{k^2},
\label{eq:enstdef}
\end{equation}
where $\bar{k^2}$ is the spectral-weighted mean of $k^2$.
The enstrophy can be obtained directly from the simulation data by application of the numerical, finite difference,  curl operation on
the primitive flow fields. 
A potential issue is that finite difference gradient operations on a compressible flow can pick up unphysical, numerical noise that obscures the signals of interest. Previously we referred to this as finite difference ``cross-talk''. However, \cite{pjr15} demonstrated that these effects can
be significantly ameliorated by using a simple smoothing operation on the
velocity fields (``Favre filtering''), without significantly reducing the desired signal. Consequently, we employ the same approach in
our enstrophy analysis here, employing a simple $3^3$ cell-average smoothing.

For Kolmogorov solenoidal turbulence the power spectrum can be written as $E_s(k) = C_0 \eta_d^{2/3}k^{-5/3}$ \citep[e.g.,][]{gotoh}, where $\eta_d$ is the rate of solenoidal turbulent energy
dissipation {\it per unit mass}, and $C_0 \sim 3/2~-~2$, is the so-called Kolmogorov constant.
Applying the Kolmogorov form over a 
range of wavenumbers $[k_0,k_1=a_k k_0]$, with $a_k > 1$ Eq.~\ref{eq:enstdef} leads to the relationships,
\begin{eqnarray}
\epsilon = \frac{1}{4}\langle v_{\rm sol}^2\rangle k_1^2~a_k^{-2/3}\frac{[1-a_k^{-4/3}]}{[1-a_k^{-2/3} ]}\nonumber\\
=\frac{3}{4} C_0 \eta_d^{2/3} k_1^{4/3}\left[ 1 - a_k^{-4/3}\right].
\label{eq:enstvel}
\end{eqnarray}

The information in Eq.~\ref{eq:enstvel} can also be used to express the turbulent
energy dissipation rate, or energy flux rate, in terms of either the solenoidal velocity or the enstrophy. In the limit $a_k \gg 1$ these become,
\begin{equation}
\eta_d \approx \left(\frac{1}{3 C_0}\right)^{3/2} \left(\langle v_{sol}^2\rangle\right)^{3/2} k_0 \sim 0.5 \frac{\left( \delta v_{L}\right)^{3}}{L},\\
\label{eq:veldis}
\end{equation}
\begin{equation}
 \approx \left(\frac{4}{3 C_0}\right)^{3/2}\frac{\epsilon^{3/2}}{k_1^2}\sim 0.014 \epsilon^{3/2} \ell_1^2,
\label{eq:ensdis}
\end{equation}
where in the final forms we have set $C_0 = 1.8$,
$\langle v_{\rm sol}^2\rangle)^{3/2}\sim \delta v_L^3$ with $k_0 = 2\pi/L$ to match the energy flux relation in equation \ref{eq:keflux} and set $a_k = L/\ell_1$.
We tested the relations Eq.~\ref{eq:veldis} and Eq.~\ref{eq:ensdis} using steady, driven, homogeneous turbulence simulation
data (of known dissipation rate) presented in \cite{pjr15} and found good agreement to within $\sim 15$ \% for both
the velocity-based and enstrophy-based predictions.
The turbulent energy dissipation rate per unit volume can be expressed as
$f_{\rm turb} = \rho\cdot\eta_d$. We will apply these relations to our cluster simulation data in Sec.~\ref{subsec:enstanal}.
In developed hydrodynamical turbulence the
total rate of that dissipation is independent of the microphysical details, although it obviously provides
upper bounds to 
energy input rates.

\begin{figure*}
\includegraphics[width=1.0\textwidth]{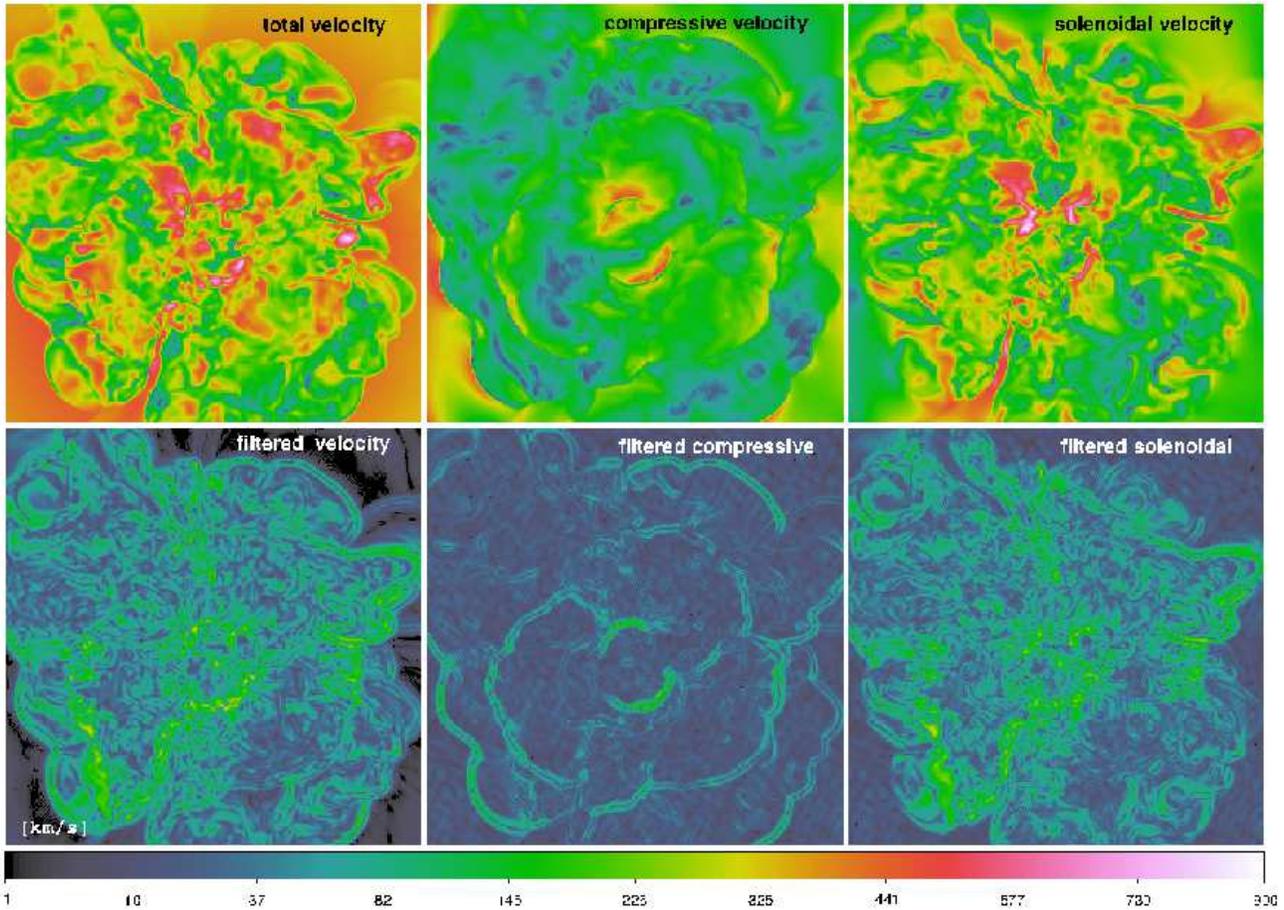}
\caption{Velocity field for the same slice as Fig.\ref{fig:cslice}, according to different filtering strategies. Top row:
magnitude of the total velocity field, compressive and solenoidal velocity components. Second row:
Magnitude of the small-scale filtered turbulent velocity (Sec.~\ref{subsec:turbo} for details), compressive and 
 solenoidal turbulent  components of the velocity field.} 
 \label{fig:vfield}
\end{figure*}

One of the keys to a useful understanding of ICM turbulence is
an understanding of how, where and when it is generated.  Enstrophy tracking provides an effective and practical tool to study this for the dominant, solenoidal component. 
To this end we can derive an equation for enstrophy evolution using the curl of the Navier-Stokes equation
\citep[e.g.,][]{pjr15}; namely,
\begin{equation}
\frac{\partial\epsilon}{\partial t}=F_{\rm adv} + F_{\rm stretch} + F_{\rm comp} + F_{\rm baroc}, +  F_{\rm diss},
\label{eq:enst}
\end{equation}
where enstrophy source, sink and flux terms, $F$, (all called ``source terms'' below) are defined as,
\begin{eqnarray}\label{eq:enstsource}
F_{\rm adv} =- \nabla\cdot (\vec v \epsilon) =-( \epsilon\nabla\cdot\vec v + \vec v\cdot\nabla\epsilon),\nonumber\\
F_{\rm stretch} = \vec\omega\cdot (\vec\omega\cdot\nabla)\vec v =  2\epsilon (\hat\omega\cdot\nabla)\vec v \cdot \hat\omega,\nonumber\\
F_{\rm comp} = -\epsilon \nabla\cdot \vec v = \frac{\epsilon}{\rho}\frac{d\rho}{dt} = -\nabla\cdot (\vec v \epsilon) + \vec v\cdot\nabla\epsilon,\\
F_{\rm baroc} = \frac{\vec\omega}{\rho^2} \cdot (\nabla\rho\times\nabla P)\nonumber,\\
F_{\rm diss} = \nu\vec\omega\cdot\left(\nabla^2\vec\omega + \nabla\times \vec G\right).\nonumber
\end{eqnarray}

The enstrophy advection term, $F_{\rm adv}$, in Eq.~\ref{eq:enstsource} is conservative, so that its integral over a closed system must vanish. We will see over cluster volumes, however, that the integral of this term does not vanish. The $F_{\rm stretch}$, $F_{\rm baroc}$ and $F_{\rm comp}$ terms account for vortex stretching, enstrophy production in  baroclinic flows and in compressions, respectively. Note that the fluid compression rate, $-\nabla\cdot\vec v$, enters into both the $F_{\rm adv}$ and $F_{\rm comp}$ terms. However, whereas $F_{\rm adv}$ always integrates to zero in a closed system, $F_{\rm comp}$ does not if there is a net alignment of the velocity with the enstrophy gradient field. This alignment is usually present in shocks, so that  $\int F_{\rm comp}~dV = \int \vec v\cdot\nabla\epsilon~dV > 0$ there, but is mostly small elsewhere in the absence of systematic compression \citep{pjr15}. In driven turbulence ``in a box'' simulations this term was found to provide a good, overall measure of enstrophy growth by way of shocks \citep{pjr15}. During cluster mergers there will be systematic compressions and rarefactions, and the behavior of the $F_{\rm comp}$ term will reflect that, as well. True vorticity source terms such as vorticity creation in curved or intersecting shocks, or $F_{\rm baroc}${\footnote{Curved or intersecting shocks can create vorticity even in isothermal flows, whereas $F_{\rm baroc}$ cannot. So, these sources represent distinct physics.}, for example, while necessary to seed enstrophy in an irrotational flow, are mostly sub-dominant in homogeneous turbulence simulations once any vorticity exists in the flow.  We will examine the varied roles of each of these terms in our cluster simulation data in Sec.~\ref{subsec:enstanal}.

For completeness we include in Eq.~\ref{eq:enst} and Eq.~\ref{eq:enstsource} the explicit viscous dissipation term,  $F_{\rm diss}$,  where $\vec G = (1/\rho) \nabla \rho \cdot \vec {\vec S}$, with $\vec {\vec S}$ the traceless strain tensor \citep[e.g.][]{pjr15}.
However, our simulations are based on Euler-limit hydrodynamics, where there is no explicit viscosity, $\nu$. Thus it is not possible to evaluate $F_{\rm diss}$  explicitly in our simulations.  On the other hand, effective net turbulence dissipation rates can be estimated using the turbulence relations in Eq. ~\ref{eq:veldis} and  Eq.~\ref{eq:ensdis}, which we do in Sec.~\ref{subsec:enstanal}. 
See \citet[][]{2010ApJ...712....1Z,sc14} for previous analogous turbulence dissipation analyses in clusters.
We will examine this issue more broadly for the ISC simulations in a subsequent paper. 


\section{Results}
\label{sec:results}

As mentioned in Sec.~2, we focus this paper on the one cluster designated it903, while we defer the study of the complete ISC sample to future work. 


\subsection{Preliminary turbulence analysis.}
\label{subsec:results_maps}

We start our analysis of turbulence in it903 by studying the spatial distribution of the gas velocity field,
filtered according to the methods presented in Sec.~\ref{subsec:turbo}-\ref{subsec:helm}.
 
Fig.~\ref{fig:vfield} illustrates a 2D slice of the velocity field at $z= 0.32$, processed
in several ways to reveal its turbulence properties: unfiltered total velocity along with its compressive and solenoidal
components (top row) or small-scale filtered velocity field (as in Sec.~\ref{subsec:turbo}) and its components (bottom row). The distribution of filtering scales used to remove large-scale motions has been shown in the bottom panel of Fig. \ref{fig:fkin_scales}.

The general evolution of this cluster,  until the last major merger close to $z \approx 0.3$, can be seen in the sequences
of 3D volume-rendered images in Fig.~\ref{fig:vrender}.
The merger developed along the upper-left, lower-right diagonal of this view.  While the volume distribution of shocks visible in Fig. \ref{fig:vrender}  is quite complex, the 2D slice in Fig. \ref{fig:cslice} reveals two fairly clear merger shocks near the centre of the shock image and about 500 kpc from the cluster centre at this epoch. The estimated Mach numbers are $\mathcal{M} \sim 2.5-3$ in each case. Stronger shocks are 
visible at larger distances in both the 3D and 2D images. Mach numbers approach $\mathcal{M} \sim 10^2$ for  outer accretion shocks. While the accretion pattern of this object is dominated by the large-scale filamentary accretion along the merger axis, small filamentary accretion patterns are detected in other directions. 

Before entering through accretion shocks, the accreted gas reaches typical infall velocities of 
 $\sim 500-700~\rm km/s$  at this epoch. This flow is predominantly compressive, yet significant
 solenoidal velocity components are present even before crossing accretion shocks, where filaments break the spherical geometry of accretion shocks (e.g., in the top right corner of the image). Shock interactions also significantly enhance the solenoidal motions (see Sec. \ref{sec:howwhere}).
 
Inside the cluster the coherence of infall motions gets broken 
by irregular,  converging flows and resulting shear motions. The maxima in the velocity field, associated with the density peaks of substructures, can reach $\sim 500-800 \rm ~km/s$. 
Even the unfiltered velocity field gives a clear visual impression of a predominance of solenoidal
motions within the cluster. The principal exceptions are associated with strong shock waves sweeping through the volume at all times, but especially during the major merger.

The actual predominance by solenoidal turbulent motions is clearly revealed when large-scale laminar motions and shocks
are filtered out (lower right panel of Fig.\ref{fig:vfield}). The total filtered
velocity (lower left panel) is dominated by solenoidal motions everywhere except near shocks, where small-scale velocity structures ($\sim 100 ~\rm km/s$) are impossible to distinguish between real small-scale turbulence and simpler shock jumps. Everywhere at distances $\geq 100 ~\rm kpc$ away from shocks  the small-scale velocities are almost entirely solenoidal. The absolute maxima of the small-scale solenoidal velocity field are found  at the interface of filamentary accretions within the cluster volume, and also downstream of shocks. We will analyse the generation/amplification of vorticity at shocks in the next section.

 \begin{figure}
\includegraphics[width=0.45\textwidth]{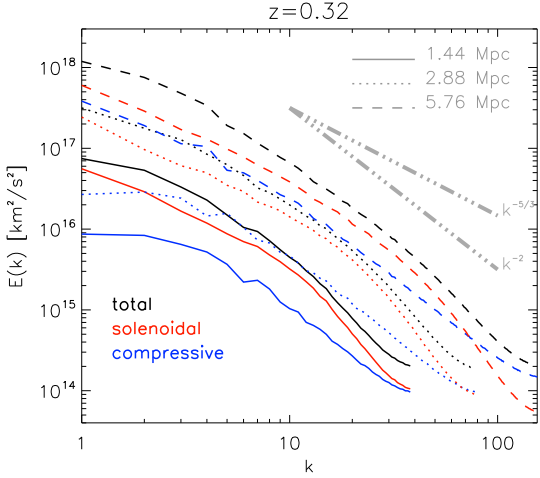}
\caption{Power spectrum of the 3D velocity field within increasing volumes around the centre of cluster it903 at $z=0.32$. The black lines show the spectra for the total (unfiltered) velocity, the red (blue) lines show the spectra of the solenoidal (compressive)  velocity components. The additional straight grey lines show the $\propto k^{-5/3}$ and $\propto k^{-2}$ for comparison. In each line, the wavenumber $k$ is referred to each specific  volume (i.e.  $k=1$ references $1.44$, $2.88$ or $5.76$ Mpc. For clarity the spectra of different boxes have been multiplied by the corresponding volumes.} 
 \label{fig:pk}
 \end{figure}

 \begin{figure}
\includegraphics[width=0.45\textwidth]{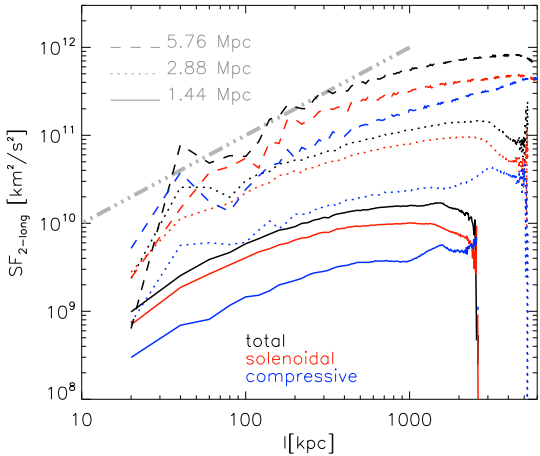}
\caption{Second-order longitudinal structure functions for the same regions
of Fig.\ref{fig:pk}, with identical meaning of colors and linestyles. The additional straight grey line shows the $S_{2}(l) \propto l$ trend for comparison. For clarity the structure functions of different boxes have been multiplied by the corresponding volumes.}
 \label{fig:sf}
 \end{figure}
 
 \begin{figure}
\includegraphics[width=0.45\textwidth]{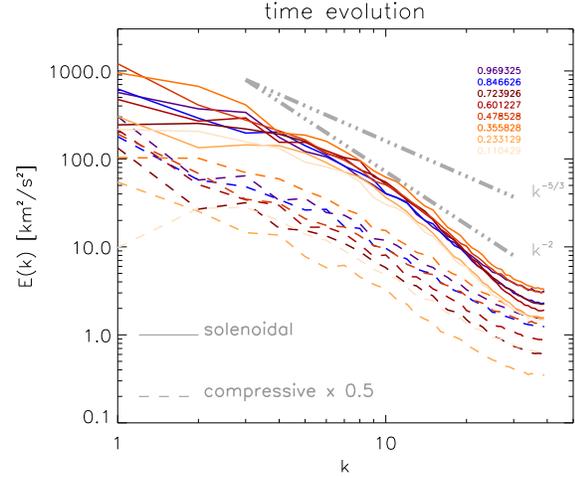}
\includegraphics[width=0.45\textwidth]{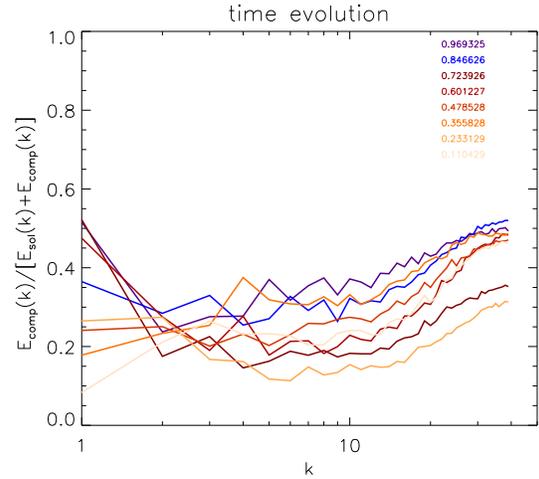}
\caption{ Top panel: evolution of the power spectrum of the solenoidal and compressive velocities for the inner, cluster-centred $(1.44 \rm ~Mpc)^3$ region of it903. The compressive component has been rescaled by a factor $0.5$ for a better visibility of all curves. Bottom panel: ratio of the compressive to the total velocity power spectrum as a function of wavenumber, for the same redshifts of the top panel. The wavenumber $k$ has the same scale of  Fig.\ref{fig:pk}, for the $(1.44 ~\rm Mpc)^3$ volume.}
 \label{fig:spec_evolve}
 \end{figure}

\begin{figure}
\includegraphics[width=0.45\textwidth]{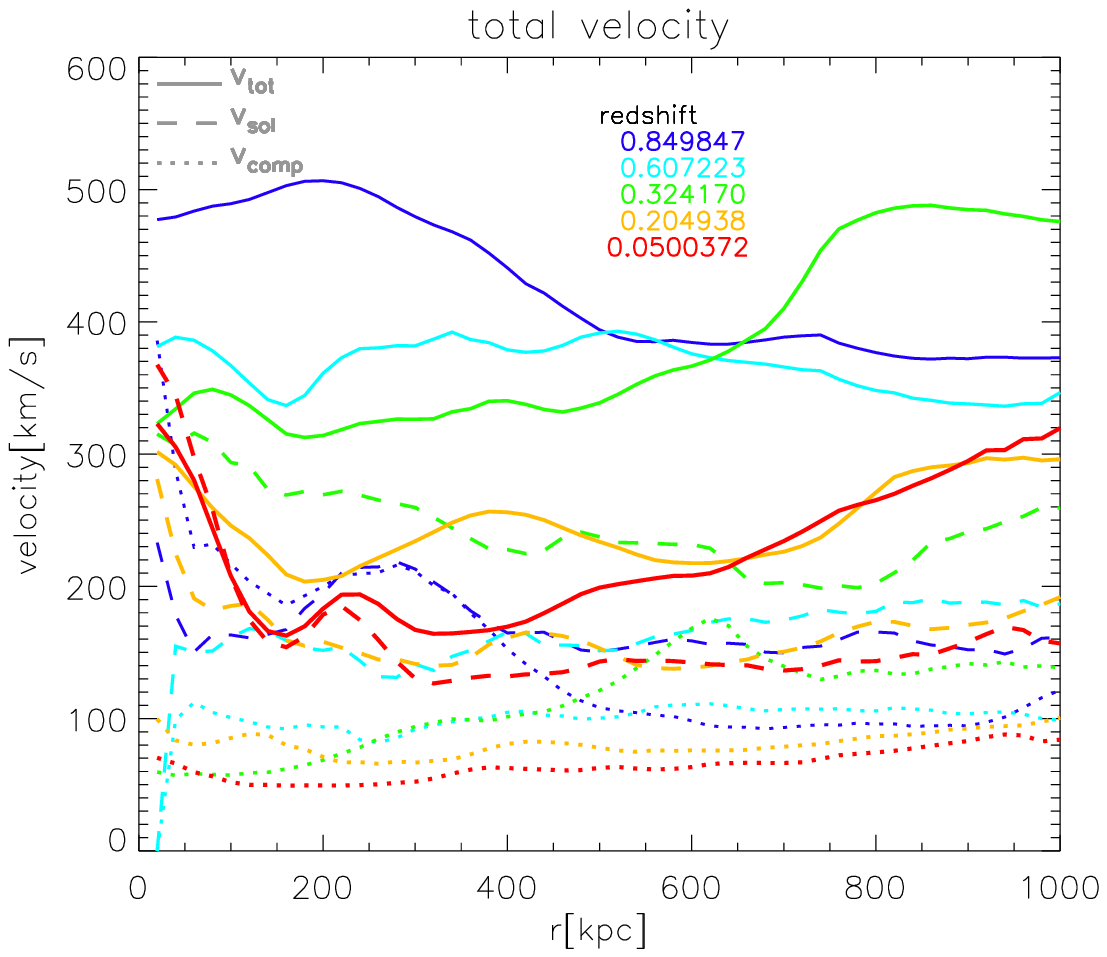}
\includegraphics[width=0.45\textwidth]{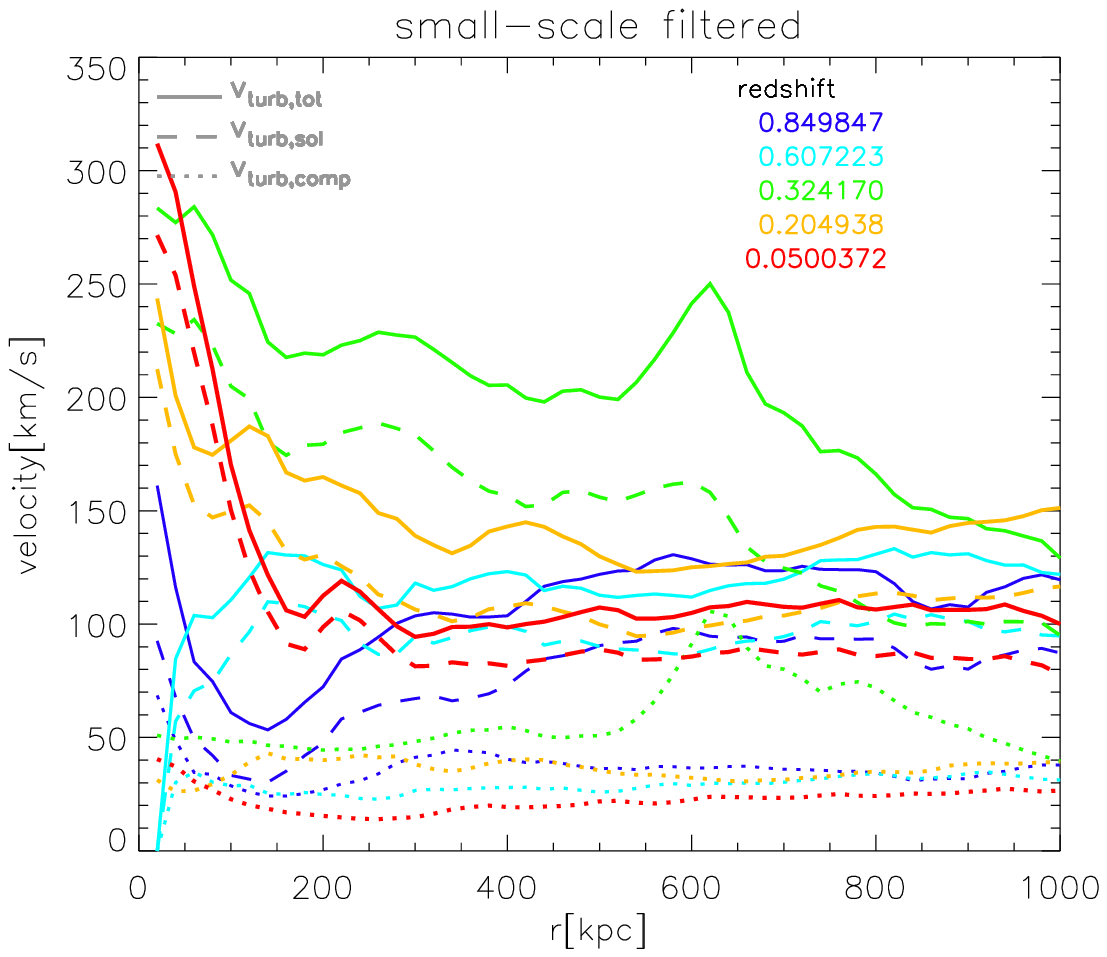}
\caption{Average radial velocity profiles for it903 at different redshifts.
Top panel: average total (unfiltered) velocity profile (solid), further decomposed into solenoidal (dashed) and compressive (dotted) components. Bottom panel: average small-scale filtered velocity profiles for the same redshifts, according to our procedure of Sec.\ref{subsec:turbo} with masking of shocks with $\mathcal{M} \geq 1.2$.}
 \label{fig:prof_vel}
\end{figure}

We next analyse the turbulent nature of the flow using power spectra of the (unfiltered) velocity field shown in Fig.~\ref{fig:pk}. For that we computed the 3-D power spectra within increasing volumes around the centre of it903 at $z=0.32$, assuming, for that exercise, periodic boundaries in application of FFTs. The spectra for the total velocity field show the typical power-law behaviour of clusters simulated in this way \citep[e.g.,][]{va11turbo}, up to nearly 
two orders of magnitude in scale when the cluster virial volume$(5.76 ~\rm Mpc)^3$ is considered. The spectrum flattens at low k-values, but the power-law is close to $E(k) \propto k^{-2}$ (i.e. steeper than Kolmogorov turbulence) for most scales. 
Fig.~\ref{fig:sf} shows the complementary view of the second-order longitudinal structure function for the same volumes, obtained by randomly extracting $\approx 5 \cdot 10^{4}$ paris of cells in the domain. The trend of the structure functions
is similar to the results of \citet[][]{miniati14}, with hints of a flattening at $\sim \rm ~Mpc$ scales and of a steeper behaviour of the compressive
component at small scales.
In both panels we also show in colors the spectra/structure functions of the solenoidal and compressive components. Again, this shows how the solenoidal component is larger at most scales. However, in the largest box the difference between the two modes is reduced, and in this case the smallest scales are dominated by the compressive component, suggesting the relevant contribution of shocks forming in cluster
outskirts.  When larger volume are included, the compressive structure functions steepen at small scales, strengthening the
view that shocked regions become increasingly more relevant on large scales.
In both views the recovered trends are consistent overall with the picture of  a turbulent ICM, mixed with large-scale regular velocity components
for scales $\geq 0.1-1~\rm Mpc$ and punctuated by small-scale velocity perturbations due to shocks. 

We notice that our analysis detects a significantly steeper slope (both in the power spectra and in the structure function) in the solenoidal component, compared to the compressive component. This is at variance with some other recent numerical studies of turbulence in the ICM \citep[e.g.][]{pjr15,miniati15}. However, understanding the origin of this difference is not trivial. Most of the difference is seen at small scales, when the impact of numerical dissipation is larger \citep[e.g.][]{kri11}. Moreover, the mass/dynamical state of it903 is different from the one analysed in \citet{miniati15}, and also our method for the mode decomposition of turbulent modes is different (Sec.~\ref{subsec:helm}, see also Appendix A.1.2). Constraining the slope of turbulent modes at small scales in the ICM is relevant to estimate of (re)acceleration of radio emitting particles \citep[e.g.][]{bj14,miniati15,br16},  but we defer a more extensive exploration of this issue to future work with our ISC sample.

\begin{figure*}
\includegraphics[width=0.95\textwidth]{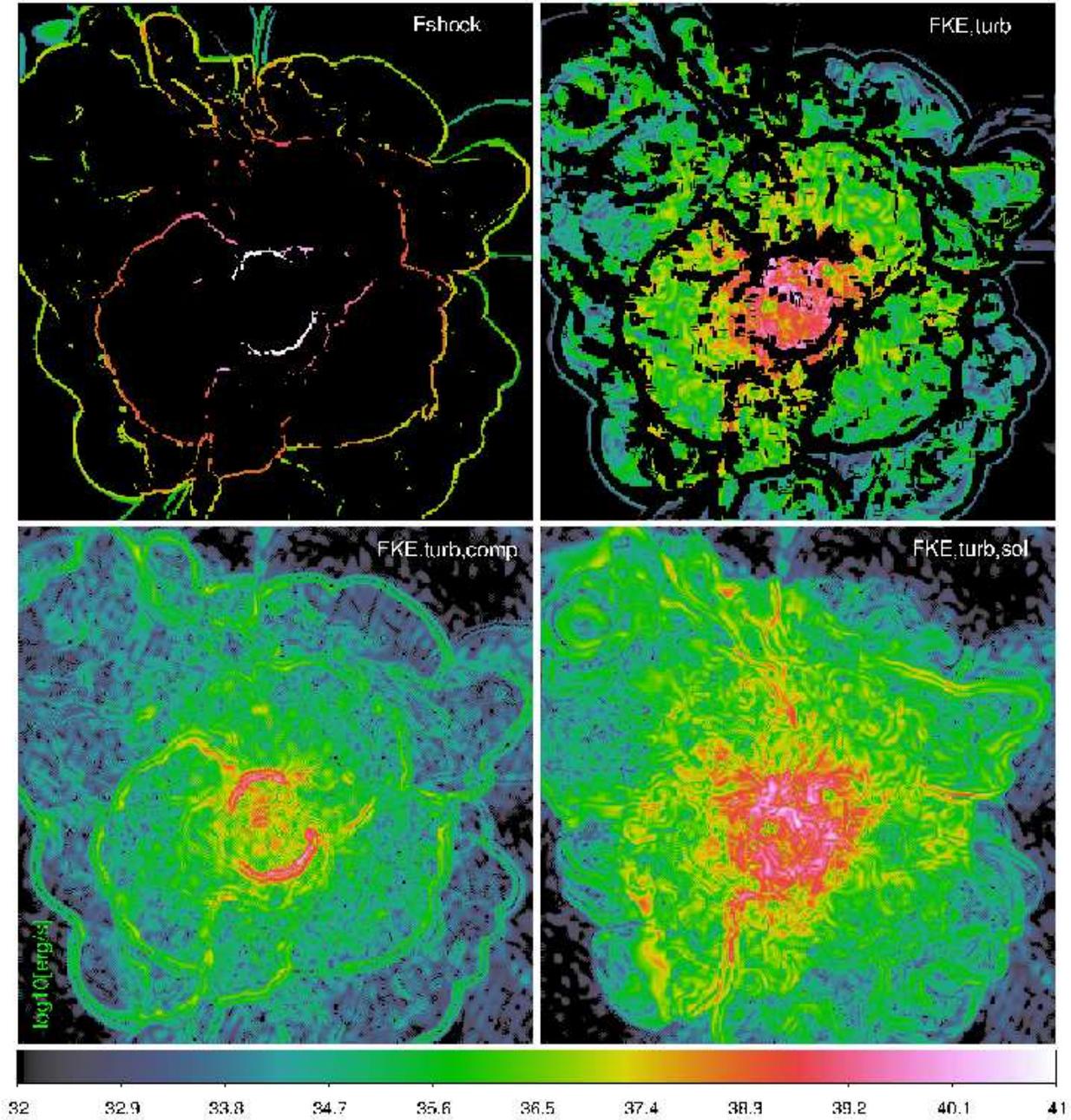}
\caption{Distribution of shock kinetic power (top left), filtered compressive (bottom left) and filtered solenoidal (bottom right) kinetic energy flux for the same slice of Fig.\ref{fig:vfield}. The top right panel additionally show the kinetic energy flux of the total filtered velocity field, with the additional masking of shocked regions.} 
 \label{fig:ske}
\end{figure*}

In Fig.~\ref{fig:spec_evolve} we show the evolution of the power spectrum in the inner, cluster-centred  $(1.44 ~ \rm Mpc)^3$ region of the two components at different epochs (top panel), and the ratio between the compressive and the total velocity power spectrum for the same epochs (lower panel). While the solenoidal component only  shows variations within a factor 2 for $k \geq 2$ at all epochs, the compressive component varies more significantly over time and at all scales. 
Consequently,  the ratio between the compressive and the total power spectrum also shows 
significant variations: in the investigated interval it ranges
from $\sim 5\%$ to $\sim 50 \%$ at the largest scales, and from $\sim 30\%$ to $\sim 50\%$
at the smallest scales.
However,  it is worth noticing that that the actual difference between solenoidal and compressive turbulence is not properly captured by this simple ratio. In reality large-scale velocity fields introduce regular components at all spatial scale, which are best removed only by our multi-scale filter. Likewise, shocks very significantly bias the estimate of the real compressive turbulence at the smallest scales. It is not simple to evaluate the proper role of such shocks. Some, especially weak shocks with large curvature, are truly components of the compressive turbulence (i.e., uncorrelated flows), while stronger shocks with small curvature are not. In either case said shocks can become sources of turbulence and our approach is to try to allow for a range of possibilities.

The average (volume-weighted) radial velocity profiles for different epochs of it903 are given in Fig.~\ref{fig:prof_vel}, and can be compared with the small-scale filtered profiles (bottom panel). We also show the average volume-weighted profiles of the compressive and solenoidal components. In the early stages of cluster formation the total velocity was large in the centre, $\sim 500 \rm ~km/s$. At later stages, after the major merger, it flattened at all radii and decreased to $\sim 200 \rm~km/s$ in the central regions, slightly increasing outwards. The unfiltered compressive velocity field is found to be larger than the solenodial field only in the centre of it903 at high redshift ($z=0.84$), following supersonic bulk motions associated with fast infalling gas substructures (Fig.~\ref{fig:vrender}, top panel),  while it is always smaller later on. 
The measured small-scale velocities (referred to within a $\sim 200$ kpc scale in the shock limiter case, or $\sim 400$ kpc in the case without masking of shocks, as in Fig.~\ref{fig:fkin_scales}) using our filtering approach are of the order of $\sim 100-200  \rm~km/s$ at most epochs, with a very flat profile outside the cluster core.  The small-scale compressive velocity component is found to be significant (but still smaller than the solenoidal one) close to the major merger
event at $z=0.32$. In particular, outside $ r_{\rm vir}$ we measure a very significant jump in the compressive small-scale velocity, $\times 2-3$ larger than the increase in the solenoidal component at the same radius.

An important point to stress here is that neither of these two velocity profiles characterizes the full turbulent (uncorrelated) velocity field of the ICM. For the first, unfiltered case the contribution from laminar infall motions (clearly visible in Fig.\ref{fig:vfield}) biases the velocities high compared to random components, while in the second case our filtering procedure computes the r.m.s. random velocities only within the $L_n$ reached before the filtering algorithm stops when finding a shock or converges on an average velocity within the scale, $L_n$. The scales $L_n$ will generally underestimate the true outer scale for uncorrelated motions. Moreover, the inhomogeneity of the cluster limits the meaning of these scales, as r.m.s. velocity extracted using different $L_n$ are not easily compared.

As we already commented in Sec.~\ref{subsec:turbo}, 
the kinetic energy flux represents a more robust tool than velocity magnitudes  to 
measure the consequences of random velocity components, because it is a relatively scale-independent
measure across the turbulent cascade. {\footnote{We remark that this  scale invariance is strictly valid only in the Kolmogorov regime. It is approximate here, since the spectra for the two components in this work show some degree of departure from this.}} In addition, the kinetic energy flux itself has 
important physical meaning. It bounds the dissipation rate of kinetic energy of gas motions into
thermal energy \citep[e.g.,][]{zu16a,zh16} and into cosmic ray energy \citep[e.g.,][]{bj14,miniati15}. That energy flux also feeds the amplification of 
magnetic fields via small-scale dynamo action \citep[e.g.,][]{pjr15}, although on smaller scales than we simulate here \citep[e.g.,][]{bm15}. {\it Therefore, in the remainder of this paper we will use the turbulent energy flux as our primary turbulence metric, rather than the velocity field or the kinetic energy to describe the  turbulence in it903.}

The energy flux across shocks provides an additional, specific and important ICM dynamical metric, since some fraction of this energy is dissipated into heat, while some it also feeds the generation of turbulence, as outlined above. In Fig.~\ref{fig:ske} (top left) we show the kinetic energy flux though shocks  ($f_{\rm KE,shock}$,  Eq.~\ref{eq:fcr})
and, for comparison, the kinetic energy flux of solenoidal (bottom right) and compressive (bottom left) filtered velocity fields ($f_{\rm KE,turb}$,  Eq.~\ref{eq:keflux}) for the same slice as in Fig.~\ref{fig:vfield}. To better highlight the role played by shocks, in the same Figure (top right panel) we also show the kinetic energy flux of the total filtered velocity field, after masking the region tagged as shocked (Sec.~\ref{subsubsec:shocks}).

The kinetic energy flux in the cluster  is dominated by central,
merger shocks, which process $\sim 10^{40}-10^{41} \rm ~erg/s$ per cell. However, in the innermost $(1 \rm ~Mpc)^3$  cluster volume,  downstream of the expanding merger shocks, the kinetic energy flux in the solenoidal component displays many large patches of 
high dissipation rate, with values of order $\sim 10^{39}-10^{40} \rm ~erg/s$. 

\begin{figure}
\includegraphics[width=0.45\textwidth]{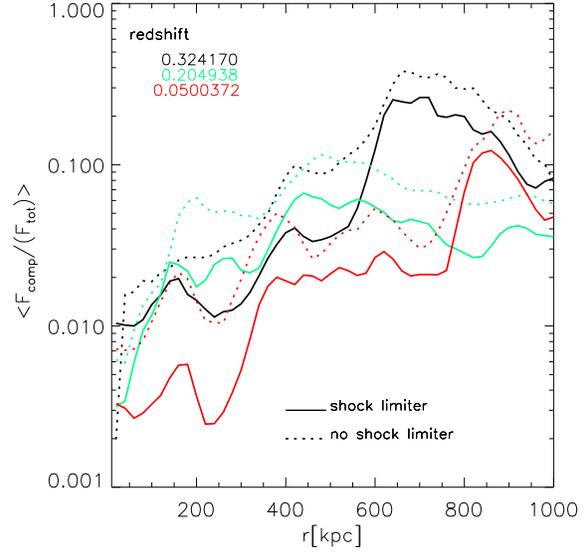}
\caption{Radial profile of the ratio between the kinetic energy flux of the small-scale compressive components and the total small-scale kinetic energy flux ((with or without our shock masking procedure in both cases). The profile are drawn for the three epochs of Fig.\ref{fig:fkin_scales}.} 
 \label{fig:profile_Fratio}
 \end{figure}

The radial profiles of the ratio of compressive to total kinetic energy flux, $f_{\rm KE,turb.comp}/f_{\rm KE,turb}$, are given in Fig.~\ref{fig:profile_Fratio}, both for the small-scale filtering procedure without masking shocks, and for the filtering procedure including  shock masking when $\mathcal{M} \geq 1.2$ (Sec.~\ref{subsubsec:shocks}).  At all epochs, the flux ratio displays a marked increase with radius.  
With shocks included, the flux ratio ratio is $\sim 2-3$ times larger (i.e. the relative energy flux of the compressive components is increased). 
When shocks are not included, the flux ratio is only $\sim$ a few percent in the central $\rm Mpc^3$volume ($\leq 0.5 ~\rm r_{\rm vir}$) at all investigated epochs. This further justifies our use of enstrophy as a trustworthy proxy of turbulence in the following Section.
 Interestingly, close to the major merger epoch ($z=0.32$ in the Figure) and at larger radii, the flux ratio jumps to $\sim 30 \%$ ($\sim 40 \%$ if shocks are included), highlightling the significant generation of compressive turbulence triggered by the merger.
We remark that, in order to better generalise this results, the analysis of a more extended set of clusters is necessary.}


\subsection{Enstrophy Analysis}
\label{subsec:enstanal}

In Sec.~\ref{subsec:vorticity} we outlined properties of fluid enstrophy, $\epsilon=(1/2) \omega^2$, that can be used effectively and efficiently to probe properties
of solenoidal turbulence. Here we apply those tools to the simulation data for the
it903 cluster. 

\begin{figure}
\includegraphics[width=0.49\textwidth]{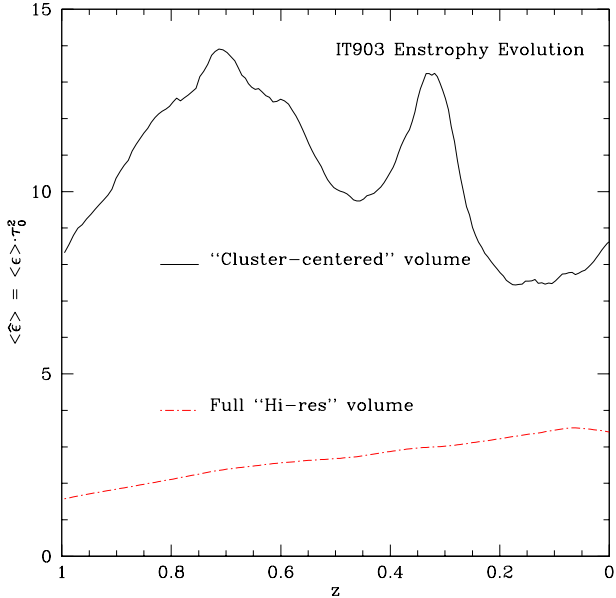}
\caption{Evolution of  volume-averaged enstrophy, $\langle\hat\epsilon\rangle$, in the it903 cluster. The black, solid curve includes only the (1.44 Mpc)$^3$ cluster-centred volume, while the red, dot-dashed curve uses the full (6.3 Mpc)$^3$ high resolution volume.  Enstrophy is normalized to the representative timescale, $\tau_0 = 1$ Gyr.}
\label{fig:it903enstevolve}
\end{figure}

Fig. \ref{fig:it903enstevolve} illustrates the
evolution since $z = 1$ of volume-averaged enstrophy within the full $(6.3 ~{\rm Mpc})^3$ high-resolution volume (dot-dashed line) and within the smaller, cluster centered, $(1.44~ {\rm Mpc})^3$ volume (solid line).  
It is obvious that signatures of the cluster dynamical history are much more evident if one focuses on a relatively small volume more closer to the virial size of the cluster.  The $z \sim 0.3$ major merger is quite obvious in the smaller volume, but not evident at all in the larger one.
The enstrophy is expressed  again as 
the normalized quantity, $\hat\epsilon = \epsilon\cdot\tau_0^2$, where the
characteristic time, $\tau_0$, with  $\tau_0 = 1$ Gyr. Then, $\sqrt{\hat\epsilon}$
represents a representative number of turbulent eddy turnovers per Gyr. 
For comparison, a characteristic eddy velocity, $\delta v_{\rm sol} \sim 100$ km/sec
and a characteristic coherence scale $L \sim 100$ kpc, lead to $1/\sqrt{\epsilon}\approx 1$ Gyr.
  Here in the smaller volume 
the characteristic $\langle\hat\epsilon\rangle\sim 10$, implying eddy turn-over times $\sim 300$ Myr.  

In addition,  the 
mean enstrophy in the smaller, cluster-centred volume is several times larger than in the bigger volume. Although the mean enstrophy evolution in the larger volume does not reveal distinct events, it does show a slow, monotonic
increase over time by roughly a factor of two, thus reflecting a gradual increase in turbulent energy per unit mass over time. 
The absence of clear signals for
discrete events in this larger volume is due to the strong cluster concentration of the enstrophy evident in Fig.~\ref{fig:vrender}, or,
analogously, concentration of the turbulent solenoidal velocity field shown in Fig.~\ref{fig:vfield} or Fig.~\ref{fig:ske}. 


\subsubsection{Comparison with solenoidal turbulent velocity field}

One of our objectives in this discussion is to establish the degree of concordance  in
our simulations
between enstrophy as outlined in section \ref{subsec:vorticity} and solenoidal turbulent velocity metrics as determined using methods outlined in Sec.~\ref{subsec:helm}. The $\langle\epsilon\rangle$ values in Fig. ~\ref{fig:it903enstevolve}
provide one simple test.
In Sec.~\ref{subsec:results_maps} we found characteristic turbulent solenoidal
velocities $v_{\rm sol} \sim 80$ km/sec (see Fig.~\ref{fig:ske}). Those values lead to
$\hat\epsilon\sim 10$ (with $\tau_0 = 1$ Gyr) provided length scales, $\bar{\ell} = 2\pi/\sqrt{\bar{k^2}} \sim 100$ kpc,
which is quite consistent with the coherence scale analysis in Sec.~\ref{subsec:results_maps}.

A second valuable example of cross-comparison between the enstrophy and velocity analysis of turbulence 
comes through evaluation of the energy dissipation rate of the solenoidal turbulence, which we expressed in terms
of the solenoidal turbulence velocity in Eq.~\ref{eq:veldis} and in terms of
the enstrophy in Eq.~\ref{eq:ensdis}. 
In Fig.~\ref{fig:compare} we illustrate the spatial distribution in a 2D slice of the
turbulence dissipation rate per volume at $z = 0.32$ from, on the left, the solenoidal turbulence (filtered) velocity
field  itself (Eq.~\ref{eq:veldis} times $\rho$), and, on the right, the enstrophy field (Eq.~\ref{eq:ensdis} times $\rho$). 
There are minor difference in the details, but, on the whole the agreement is remarkably good. 

In Fig.~\ref{fig:soldis} we provide a comparison of the volume-integrated turbulence dissipation rate
over time inside the (1.44 Mpc)$^3$ ``cluster-centred'' volume computed using the two formulations, again with
$\delta v_L$ and $L$ derived from the filtering analysis in Sec.~\ref{subsec:results_maps}  and with $\ell_1 = \Delta x$. For the most part the two dissipation rate estimates agree to much better than a factor of two. 
The good agreement between these two independent estimates of the turbulent dissipation rate (also applying information on very different spatial scales) stresses once more that we are capturing  reasonably well a turbulent-like cascade in its inertial range.

The only significant exception to the match between the two energy dissipation rates occurs early on, around $z \sim 1$. At this epoch our zoom volume is mostly transected  by large-scale converging motions on the proto-cluster. These motions have coherence scales of the order of the box size, which makes it impossible for our multi-scale filter to correctly disentangle bulk and turbulent component. Thus the filter identifies as turbulence even large-scale shear motions outside of the proto-cluster, which did not have enough time to cascade down to the scale where enstrophy is measured. However, this problem quickly disappears as the cluster volume grows and the "zoom" region is mostly filled by the virial cluster volume. 

\begin{figure*}
\includegraphics[width=0.95 \textwidth]{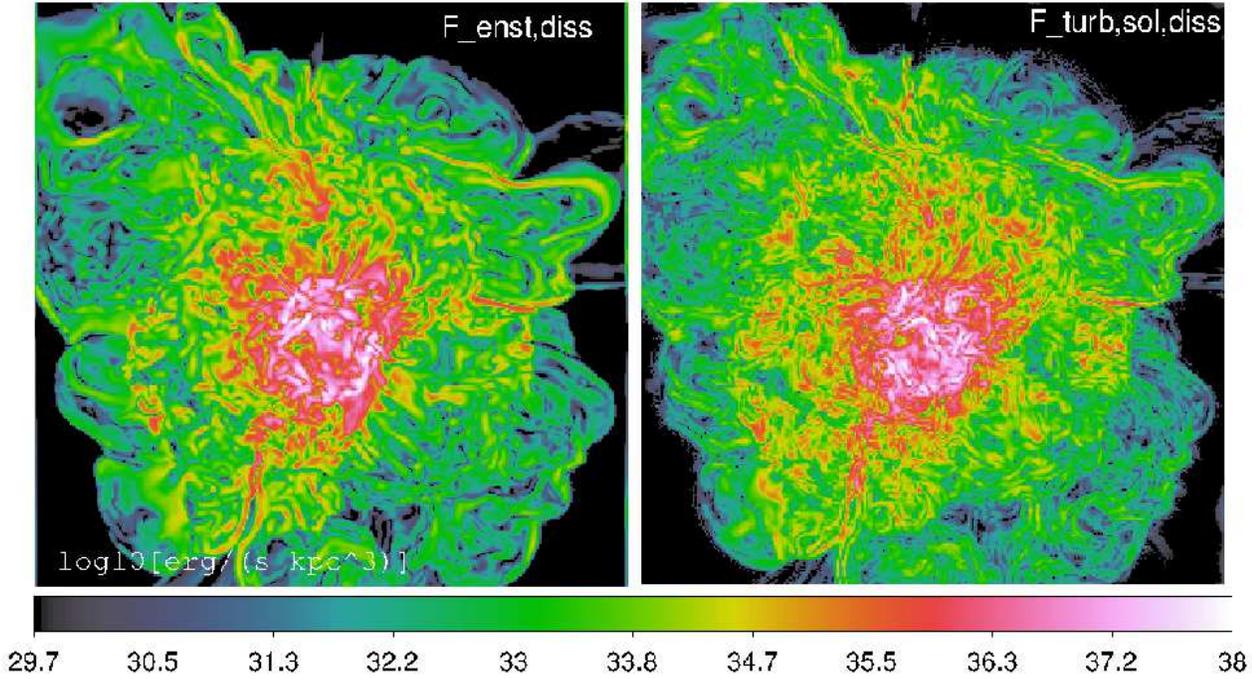}
\caption{2D slice at $z = 0.32$ (same as in Fig. \ref{fig:cslice}) showing (left)
the turbulence dissipation rate per unit volume based on the solenoidal
turbulence velocity and Eq.~\ref{eq:veldis} (or Eq.~\ref{eq:keflux}) and (right)
the equivalent turbulence dissipation rate based on the enstrophy and Eq.~\ref{eq:ensdis}.}
\label{fig:compare}
\end{figure*}

\begin{figure}
\includegraphics[width=0.49\textwidth]{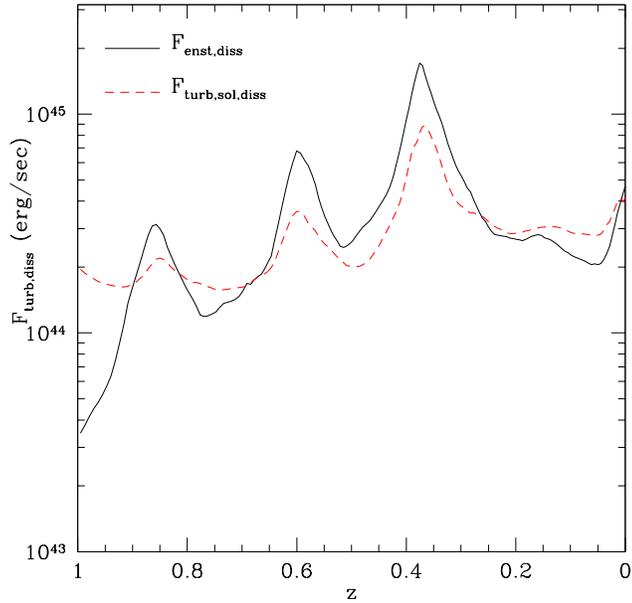}
\caption{Turbulence energy dissipation rate integrated over the (1.44 Mpc)$^3$ comoving cluster-centred volume
from the (filtered) solenoidal turbulence velocity field (red dashed curve) and from the enstrophy
distribution (black solid curve).}
\label{fig:soldis}
\end{figure}

\begin{figure}
\includegraphics[width=0.49\textwidth]{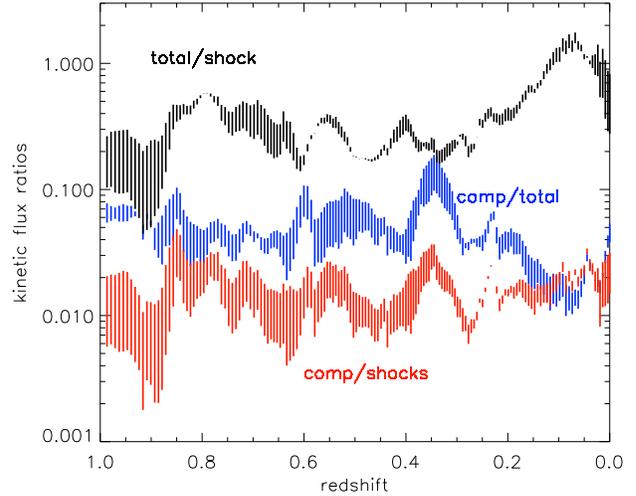}
\caption{Evolution of different energy fluxes for it903 (averaged within the central $1.44^3 \rm ~Mpc^3$ region: a)  total kinetic energy flux over shock kinetic power (black); b) compressive kinetic energy flux over total kinetic energy flux (blue); c) compressive kinetic energy flux over shock kinetic power (red). The spread in the ratios at each redshift indicates ranges related to the inclusion or exclusion of shocked regions to compute the kinetic energy flux (the upper bounds being the estimates including shocked regions).}
\label{fig:ratios_evolve}
\end{figure}

In addition to the major merger event around $z \sim 0.3$, there are other, recognizable turbulence evolution features visible in Figs. \ref{fig:it903enstevolve} and
\ref{fig:soldis}. These
include a broad enstrophy peak in Fig.  \ref{fig:it903enstevolve} between $z \sim 0.8$ and $z \sim 0.6$ with a maximum around $z \sim 0.7$. This peak breaks into a pair of peaks near $z \approx 0.85$ and $z \approx 0.6$ in the turbulent energy dissipation rate. The enstrophy evolution plot exhibits shoulders at those times, but the peak is clearly offset in time. 

Similarly, a close comparison of the enstrophy peak associated with the major merger shows that the turbulent energy dissipation peaks around $z \approx 0.32$, whereas the enstrophy itself peaks  $\sim 1/2$  Gyr later. 
Both behaviors are associated with the major merger event, but represent
somewhat different dynamics. In particular, the turbulence dissipation rate
actually peaks just before the closest approach of the two subcluster cores, when
the enstrophy is most highly concentrated into the regions of highest gas density.
The turbulent energy and its dissipation are also then focused  into these
regions. Dissipation rates outside the core regions remain relatively smaller. On the
other hand, the sharp decrease in the mean enstrophy after core passage near $z \approx 0.32$  is
actually not so much a consequence of dissipation as of the
strong outflows following the merger shocks generated during the event. In fact, as
we point out in the section below, there is a net outflux of enstrophy from this central volume. There is, in addition,  systematic decompression of the gas, which as equation \ref{eq:enst} emphasizes, leads to enstrophy reduction.

The two earlier spikes in turbulent energy dissipation evident in Fig. \ref{fig:soldis} also correspond to merging activity, although minor mergers only. In each case
there are brief intervals when turbulent motions are
concentrated into the cluster core, which leads to sharp rises in the turbulent
dissipation rate. The immediate impact on the total enstrophy budget is
less significant in these cases.

Related information is illustrated in  Fig.~\ref{fig:ratios_evolve}. It shows how ratios of various kinetic energy
fluxes evolve with time in the inner, cluster-centred volume. For each redshift we show, both, the values obtained by removing 
the contribution from shocked regions (lower bound of each color) or by including  $\mathcal{M} \geq 1.2$ shocks 
in the  turbulent kinetic flux (upper bound).  While the specific value of each ratio can change up to a factor $\sim 2$ for most of the evolution, most of the time features are seen in both
cases, and are in phase with the spikes in turbulent dissipation already noticed in Fig.\ref{fig:soldis}. In particular, each of the large spikes ($z \approx 0.85$, $\approx 0.6$ and $\approx 0.38$) is associated also with the increase of the compressive kinetic flux, which reaches $\sim 10-15\%$ of the total flux. 
Away from these spikes,  the dissipation of turbulent motions is generally contributed by solenoidal motions at the $\sim 95\%$ level. In this central volume the total kinetic energy flux is smaller than the kinetic power of shocks at most redshifts, with the exception of $z \leq 0.1$ when the two becomes comparable in the absence of significant shock waves crossing the domain.


\subsection{How, where and when is solenoidal turbulence generated in the ICM?}
\label{sec:howwhere}

The previous analysis suggests that, while the budget of purely compressive turbulence is subject to uncertainties related to the presence of shocks, 
enstrophy provides consistent measures for the local and global solenoidal turbulence.  Now we look at the processes that generate
enstrophy described in Eq.~\ref{eq:enst} and \ref{eq:enstsource}, which allows a deeper understanding of the sources and amplification of turbulence in the cluster over time. Fig.~\ref{fig:it903enstflux} shows the time evolution of the enstrophy source terms defined in the above equations. 
Analogous to the enstrophy plots in Fig.~\ref{fig:it903enstevolve}, we apply a normalization factor $\tau_0^3$. The various ratios, $\epsilon\tau_0^2/F_{\rm x}\tau_0^3$, then provide measures of the growth (or damping) timescale due to a given source term, $F_x$,  measured in time units, $\tau_0 = 1$ Gyr. Indeed, by comparing Figs.~\ref{fig:it903enstevolve} and \ref{fig:it903enstflux} we can confirm that the timescales for the various source terms in this volume are $\tau \sim~1-3$ Gyr, consistent with the apparent evolution timescale of the enstrophy in Fig.~\ref{fig:it903enstevolve}. 

As a reminder, the $F_{\rm adv}$ term, analogous to $\nabla\cdot \vec v \rho$ in the mass conservation equation, measures the net enstrophy influx rate, while
$F_{\rm stretch}$ relates to the net rate at which vortex
tubes are lengthening, $F_{\rm stretch}>0$, or shortening, $F_{\rm stretch}<0$. The $F_{comp}$ term identifies regions where enstrophy concentration correlates with ongoing gas compression. That can be reversible or not, depending on whether the gas compression is reversible or not (i.e., in shocks).  The $F_{\rm baroc}$ enstrophy source term identifies where non-vanishing cross products of density and entropy gradients align with the local vorticity (which may be expected downstream of nonplanar shock structures, for example).

We note three obvious properties of the individual source terms as  revealed in Fig. \ref{fig:it903enstflux}. The first is that all the source terms, $\langle F_{\rm adv}\rangle$, $\langle F_{\rm comp}\rangle$, $\langle F_{\rm stretch}\rangle$ and $\langle F_{\rm baroc}\rangle$ averaged over this $(1.44~{\rm Mpc}^3)$ volume are roughly comparable. During merger events around $z \sim 0.7$ and $z \sim 0.3$, when the mean enstrophy is most rapidly increasing (Fig.~\ref{fig:it903enstevolve}), the $\langle F_{\rm stretch}\rangle$ term dominates, but only at most by a factor $\sim 2$. During the major merger event around $z \sim 0.3$, there is also a sharp peak in the compressive source term, $\langle F_{\rm comp}\rangle$, in this volume. 

The second notable outcome revealed in Fig.~\ref{fig:it903enstflux} is that, despite the fact that $F_{\rm adv}$ is a conservative quantity, so that $\int F_{\rm adv} dV = 0$ over a closed volume, during most of the cluster history $\langle F_{\rm adv}\rangle > 0$ in this volume. This  highlights the fact that as a part of the accretion building the cluster, substantial enstrophy is also added into the central regions from outside.  In fact, this enstrophy accretion is, most of the time, competitive with locally generated enstrophy growth (the other three terms). Enstrophy accretion into the cluster central region is identified below as at least partly a consequence of vorticity generated near the accretion shocks outside the cluster. We note, finally in reference to Fig.~\ref{fig:it903enstflux} that there is a  brief period around $z  \sim 0.25$, following the major merger when both $\langle F_{\rm adv}\rangle $ and $\langle F_{\rm comp}\rangle$ become slightly negative in  this volume. That behavior reflects an expansion and outflow of gas from the cluster during this same interval that is visible also in Fig.~\ref{fig:fig0}. 

\begin{figure}
\includegraphics[width=0.49\textwidth]{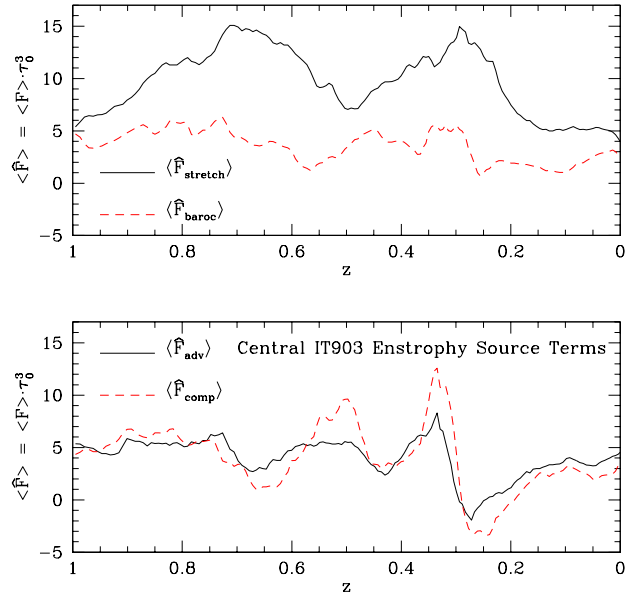}
\caption{Evolution of the source terms for enstrophy averaged over the (1.44 Mpc)$^3$ cluster-centred volume of the it903 cluster. Quantities (each $\propto~t^{-3}$) are normalized by the representative timescale, $\tau_0 = 1$ Gyr.}
\label{fig:it903enstflux}
\end{figure}

Key insights into the origins of the enstrophy within the cluster can be gained by examining the spatial distributions of the enstrophy source terms. For instance, Fig.~\ref{fig:terms} shows in the
same $6.3~{\rm Mpc}\times 6.3~{\rm Mpc}$, 2D slice used before at $z = 0.32$ (so during the major merger) the distribution of
the four enstrophy source terms. 
The distribution of the enstrophy itself in this slice (not shown) is qualitatively similar
to the distribution of filtered solenoidal velocity in the lower panel of Fig. ~\ref{fig:vfield}.
The advective source term, $F_{\rm adv}$, and especially the stretching source term, $F_{\rm stretch}$,
have very roughly similar distributions to the enstrophy, as we might expect,
since both are proportional to $\epsilon$. In detail, however,  all four distributions are quite distinct,
as we should also expect, since each depends on unique
dynamical behaviors.

\begin{figure*}
\includegraphics[width=0.75 \textwidth]{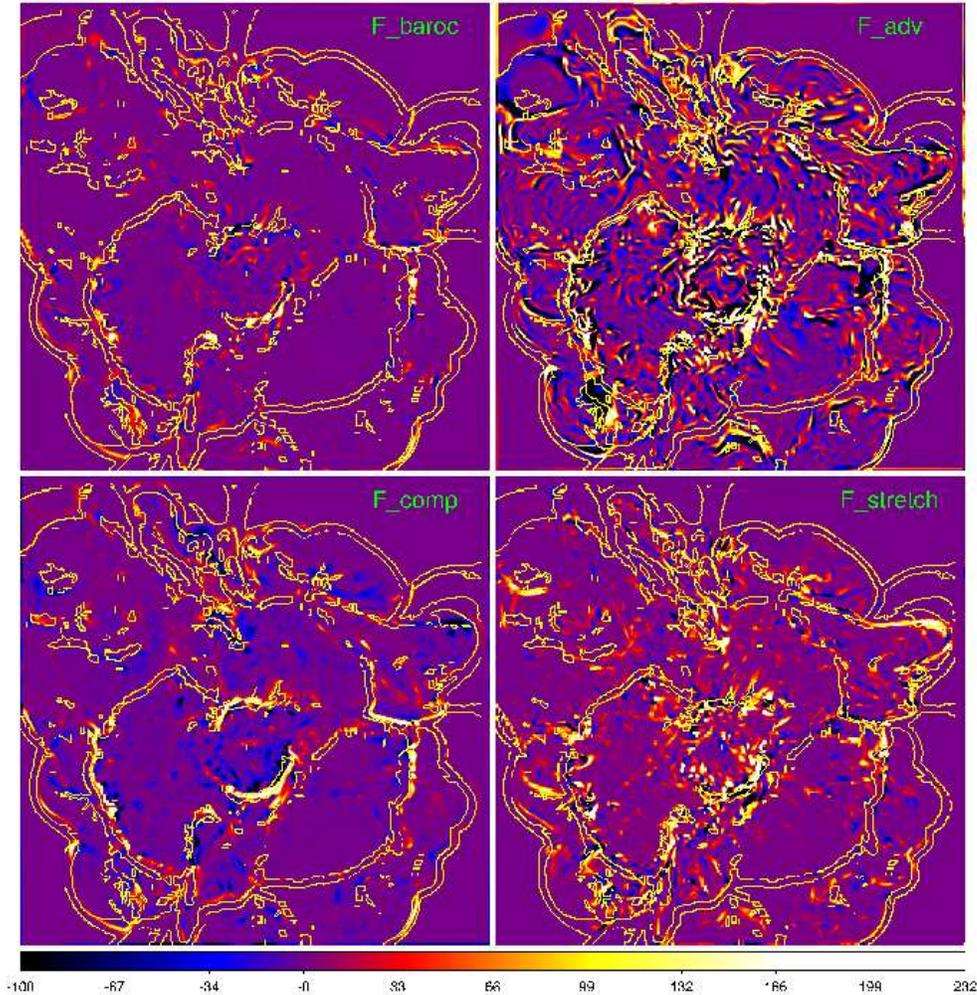}
\caption{2D distribution of enstrophy source terms in the same slice as Fig.~\ref{fig:cslice} also at z = 0.32. The source terms, defined in equation \ref{eq:enstsource}, are normalized
by the factor $\tau_0^3$, with $\tau_0 = 1$ Gyr. The yellow contours show the location of detected shocks,  based on their kinetic energy flux (only $\geq 10^{30} \rm erg/s$ cells are marked here).}
\label{fig:terms}
\end{figure*}

\begin{figure}
\includegraphics[width=0.49 \textwidth]{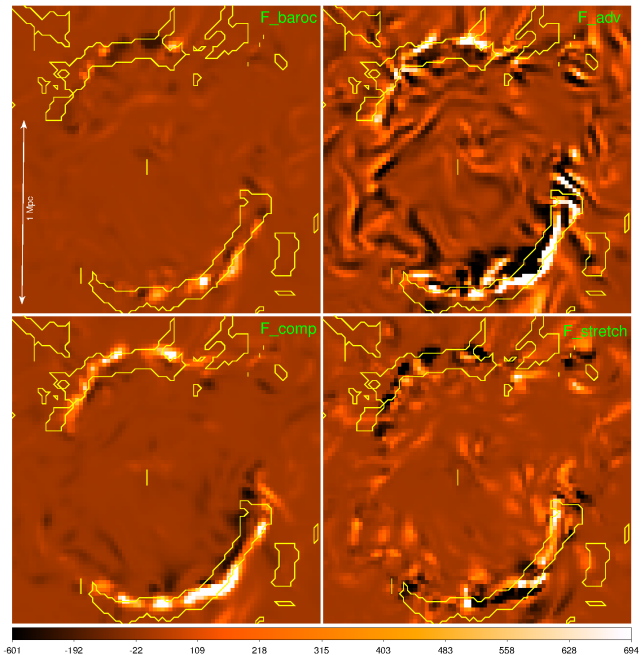}
\caption{Zoomed, $1.7~{\rm Mpc}\times 1.7~{\rm Mpc}$, central portion of Fig.~\ref{fig:terms}, focusing on the "double relic" looking shock structure that has formed as a result of the ongoing merger. The units are as in Fig.~\ref{fig:terms}. Identified shocks are again indicated by yellow contours.}
\label{fig:terms_zoom}
\end{figure}

\begin{figure}
\includegraphics[width=0.49 \textwidth]{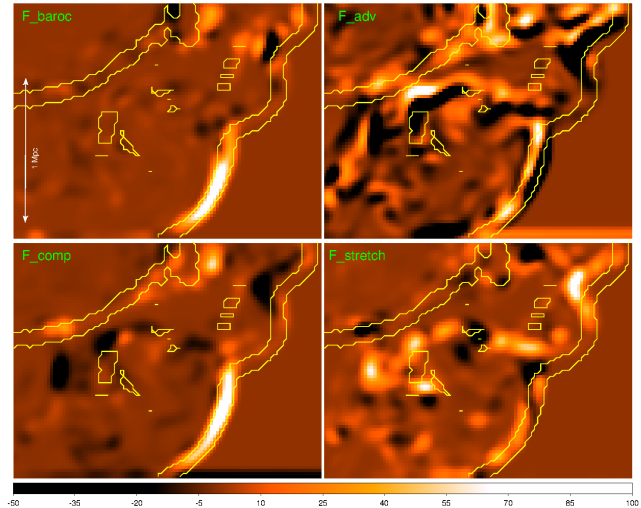}
\caption{Zoomed, $1.5~{\rm Mpc}\times 2~{\rm Mpc}$,  lower right portion of Fig.~\ref{fig:terms}, focusing on an accretion shock just above the bottom boundary.  The units are as in Fig.~\ref{fig:terms}.}
\label{fig:terms_zoom2}
\end{figure}

It is also useful to compare the source term distributions to the 
distribution of shocks in the same region (yellow contours). There is, in this context, an anticipated association between shocks and the $F_{\rm baroc}$ and $F_{\rm comp}$ terms. Indeed, \citet{pjr15}, for example, demonstrated in compressible turbulence simulations that  where the existing enstrophy is intense, {\it and} a strong shock exists, the compressive source term tends to be large. 
In this case, however, the relationship is complex.

This last point and its explanation are made clearer by examining two  zoomed $\sim (1.5~{\rm Mpc}^2)$ sub-sections of Fig. ~\ref{fig:terms}. In particular, we focus on the core cluster region between two merger shocks waves (Fig.~\ref{fig:terms_zoom}), and on the  the lower right accretion shock ($\mathcal{M} \sim 10^2$), well outside the cluster core (Fig.~\ref{fig:terms_zoom2}). Once again recall that these slices come from $z = 0.32$ during the major merger and when, according to Fig.~\ref{fig:it903enstflux}, the enstrophy growth in the cluster centred volume peaks.  In between the two internal merger shocks in Fig.~\ref{fig:terms_zoom} the $F_{\rm adv}$ and  $F_{\rm stretch}$ terms show strong, distributed patterns of enstrophy development. In comparison, the $F_{\rm baroc}$ and $F_{\rm comp}$ terms are significant only very close to the merger shocks. Both show strong positive and negative peaks. 
In contrast both the  $F_{\rm baroc}$ and $F_{\rm comp}$ terms associated with the accretion shock in Fig.~\ref{fig:terms_zoom2}  are predominantly positive, and their net contributions dominate those of $F_{\rm adv}$ and $F_{\rm stretch}$ in the area shown. 

The picture that develops from this analysis is that the vorticity (enstrophy) generation within the cluster is {\it a two step process}:
\begin{itemize}

\item Initial generation takes place in the cluster periphery in association with accretion shocks via baroclinic influences {\it and} compression. The compression contributions involve both external, accreting vortical motions and, as discussed in \citet{pjr15} creation of enstrophy through shear stresses generated within complex shock structures. Some enstrophy also is accreted, especially along filaments.

\item Subsequently, this enstrophy (solenoidal turbulence)  is accreted into the cluster ($F_{\rm adv}$), and in the innermost cluster regions  the cluster enstrophy evolves especially by stretching, $F_{\rm stretch}$, but also by compressing those vortex structures initially generated in the cluster outskirts.  Since those latter source terms are largest during merger activity, vorticity develops most strongly in those periods.  Cluster-core-scale flows and shear are dominant contributors to those drivers.

\end{itemize}

To summarize this section, we surmise, using a combination of analysis tools, that the solenoidal turbulence found in the innermost
regions of clusters is predominantly the result of injection and enhancement of accreting vortical flows at accretion shocks,  followed by significant 
amplification by central advection,  stretching and compression, particularly during merger events.The compressive and advective contributions can be at least partially reversible, whereas the stretching and baroclinic contributions generally are not.


\subsection{Comparison to previous work}
\label{subsec:comparison}

The existing literature on numerical studies of turbulence in ICMs offers several interesting comparisons to the results reported here.
Various studies have examined the distribution of global turbulent power in velocity fluctuations on different
scales, either through power spectra \citep[e.g.,][]{vbk09,va11turbo,2014A&A...569A..67G,sc14} or structure functions \citep[][]{va11turbo,miniati14,miniati15}.

The character of the ICM gas flows and the presence of turbulence in our simulated cluster is consistent with other high-resolution simulation studies in the literature \citep[e.g.,][]{do05,va11turbo,miniati14,sc14}, with quantitative differences related to the specific
analysis techniques. 
In the limited context of  high-resolution grid-based simulations of non-radiative clusters, the pressure support from turbulence in the cluster core has, for example, been estimated in the range $\sim 5-15$\% of the gas pressure using the total, global gas velocity dispersion \citep[e.g.,][]{lau09,2014ApJ...792...25N},  or, say $\sim 5-20$\% from large-scale, incoherent, but fixed scale velocity distributions, \citet{va11turbo}. On the other hand, significantly smaller turbulent pressures tend to result from multi-scale filtering algorithms (e.g., $\sim 1-5$\%, \citet{va12filter}) or from simulations employing subgrid turbulence modeling (e.g., $\sim 0.2-20$\%, \citet{2009ApJ...707...40M,sc14}). 
 
 These numbers just highlight the underlying difficulty in disentangling large-scale from small-scale motions and assessing what is the best way to extract the purely turbulent component of complex and stratified 3D flows. At that level the issue is partly one of the intended meaning of the term ``turbulence''. Is the aim, for instance, to identify an energy reservoir for heat and/or cosmic rays, or is it simply to characterize deviations from hydrostatic equilibrium? On top of that issue is the inherent uncertainty in the meaning of the velocity variations due to inhomogeneity and stratification of the cluster, the usually unsteady cluster dynamical state \citep[e.g.,][]{lau09,va11turbo}, and the 
impact of additional physics such as radiative cooling and feedback \citep[][]{va13agn,2013ApJ...777..137N,2015ApJ...806...43B} (neglected in this study). Non physical influences, including simulation numerical resolution and algorithms \citep[][]{do05,2015ApJ...806...43B} also are certain to influence these outcomes at some level. It is important to isolate as much as possible the issues, so to address them as cleanly as possible.

The closest analogy to the present study is the ``Matryoshka'' simulation by \citet{miniati14}, who, using a simulation strategy similar to ours, followed the evolution of a $\sim 10^{15} ~\rm M_{\odot}$ cluster using $\approx 10$ kpc resolution inside the virial radius (which roughly spans a $\sim 6$ times larger dynamical range) . The ``Matryoshka'' cluster gas velocity field 
was examined using structure functions and Hodge-Helmholtz decomposition to follow global evolution of solenoidal and compressive turbulence properties separately within roughly the cluster core and inside the virial radius \citep{miniati14,miniati15}.  Even though the ``Matryoshka'' cluster was $\sim 10$ times more massive than it903, the general history of the cluster and the character of evolving turbulence are roughly similar.
\citet{miniati15} found compressive turbulence to be in the range $\sim 0.2-0.4$ of the total turbulent energy in the central $(\rm Mpc)^3$ region. 

Our results are roughly similar, but suggest a significantly smaller contribution from compressive turbulence because of our filtering of shocked regions.
We found in it903 that the compressive dissipation rate is in general $\sim 5\%$ of the total dissipation rate for most of the cluster lifetime, with
a spike of $\sim 15\%$ during the major merger event (the jump in the compressive dissipation is larger in cluster outskirts, where it reaches  $\sim 30\%$ close to the merger). This translates roughly into a $15-30 \%$ energy fraction in compressive turbulence, assuming both components have identical outer scales.  Additionally, we measure a steeper spectral distribution of energy in the solenoidal velocity component than in the compressive component.

We surmise that the reduced compressive turbulence role we found is mostly due to our procedure of extracting turbulent motions (both solenoidal and compressive) from the filtered, small-scale uncorrelated
velocity field (Sec.~\ref{subsec:turbo}), rather than from the total velocity field, but also from our removal of contributions from stronger shocks to the turbulent motions on the grounds that most of those strong shocks are not directly involved in the uncorrelated motions. 

Even if the 
procedure of imposing a fixed filtering scale for the entire volume, as in \citet{miniati14}, mostly removes the large-scale
laminar velocity component \citep[see also][]{va11turbo}, that procedure cannot fully account for the variation in the 
turbulent coherence length in the stratified cluster atmosphere on intermediate scales. Our previous tests in \citet{va12filter} showed, indeed, 
that the turbulent energy budget in the cluster centre is usually reduced by a factor $\sim 2$ when a spatially varying
filtering scales is adopted.


\section{Conclusions}

Understanding the dissipation of turbulent energy is key to understand  the heating of the plasma, the acceleration of cosmic rays in the ICM, as well as the growth of intracluster magnetic fields \citep[e.g.][]{su06,bj14,miniati15}. 
 
While current X-ray line spectroscopy can provide only upper limits on the chaotic motion velocities in relatively bright cluster cores \citep[e.g.,][]{2011MNRAS.410.1797S, 2015A&A...575A..38P}\footnote{The Hitomi satellite in its short life did successfully measure velocity profiles for the Perseus cluster \citep{hitomi}}, future X-ray satellites with superior spectral resolution (e.g. {\small ATHENA}) should be able eventually to detect directly the driving-scale turbulent motions in the ICMs of multiple clusters \citep[][]{2013ApJ...777..137N,2013arXiv1306.2322E,2013MNRAS.435.3111Z,zu16b}. In the meantime,
turbulent motions in the ICM induce moderate pressure  fluctuations that may be detected in X-rays \citep[e.g.,][]{sc04,2012MNRAS.421..726S,2012MNRAS.421.1123C,2014A&A...569A..67G,2014Natur.515...85Z}, or through the S-Z effect \citep[e.g.,][]{Khatri16}.  

Numerical simulations of the ICM are fundamental to assessing the real impact of ICM turbulence on all the above. In this work, we focused on the analysis of the connection between accretion-driven shock waves and turbulent motions in the ICM. In particular, we explored both the local and the statistical causal connections between shocks and the emergence of solenoidal and compressive turbulent motions during a simulated cluster lifetime.

 Our main conclusions from this study can be summarized as follows:

\begin{itemize}

\item Gas flows in the ICM are characterised by a turbulent behaviour across a wide range of scales, roughly consistent with a Kolmogorov-like model.  However, these flows are  mixed with larger-scale regular  (correlated) velocity components for scales $\geq 0.1-1~\rm Mpc$ and are punctuated by small-scale velocity perturbations due to shocks, which makes it difficult to isolate accurately uncorrelated turbulent fluctuations of the flow at most scales (Sec.~\ref{subsec:results_maps}).

\item Using Hodge-Helmoltz decomposition within domains distributed across our simulated cluster, we measure dominant solenoidal velocity fields everywhere within the cluster and at most epochs (with the exception of high redshift epochs, when the cluster is still forming and is far from a virialised state). The solenoidal component makes $\geq 50-80 \%$ of the amplitude of the total velocity field at most epochs  and scales (Sec.~\ref{subsec:results_maps}).

\item The kinetic energy dissipation rate of the small-scale velocity field is a powerful  tool to measure the ratio of compressive and solenoidal motions in a nearly scale-independent way. The dissipation in compressive modes only accounts for a few percent of the total turbulent dissipation rate in the central $\sim \rm ~Mpc^3$ volume. This can increase to about $\sim 15\%$ in the central $\rm Mpc^3$ during major merger events, and to $\sim 30 \%$ in cluster outskirts (Sec.~\ref{subsec:results_maps}).

\item Vorticity and enstrophy are trustworthy proxies of the dominant solenoidal turbulent component.  In particular, the volume-integrated dissipation rate of solenoidal turbulence and of enstrophy are very well correlated in the $0 \leq z \leq 1$ redshift range, 
and they show remarkably similar spatial patterns (Sec.~\ref{subsec:enstanal}).

\item  For the first time, we apply the Navier-Stokes formalism to analyse in detail how enstrophy evolves in the simulated ICM, by decomposing its growth rate into advective, stretching, compressive and baroclinic terms (Sec.~\ref{subsec:enstanal}). 

\item At accretion shocks  baroclinic generation of enstrophy along with enstrophy enhancement during flow compression are the most important source terms of enstrophy. In cluster interiors vortex stretching dominates the growth of enstrophy, although advective concentration of enstrophy and, especially during mergers, enstrophy enhancement through compression can be comparable. Merger shocks  largely seed the enstrophy enhanced by vortex stretching and advective concentration in the cluster interior (Sec.~\ref{subsec:enstanal}).

\end{itemize}

The study of this first cluster of the ISC sample showed how rich is the complexity of simulated ICM turbulence, even in this rather restricted physical setup.  Our analysis suggests that a careful combination of filtering techniques is mandatory to identify all major components of the turbulent energy budget reliably, and to give them a physical meaning as a function of scale.  Through the extensive analysis of our full ISC sample in planned follow-up work it will be possible to generalise the results obtained for this first cluster in a more robust statistical way.


\section*{acknowledgements}

Computations described in this work were performed using the {\enzo} code (http://enzo-project.org), which is the product of a collaborative effort of scientists at many universities and national laboratories. We gratefully acknowledge the {\enzo} development group for providing extremely helpful and well-maintained on-line documentation and tutorials. The reported simulations and most of the analysis were carried out using resources at the University of Minnesota Supercomputing Institute (MSI). 
FV acknowledges financial support from the grant VA 876-3/1 and partial financial support from the 
FOR1254 Research Unit of the German Science Foundation (DFG). 
TWJ acknowledges NSF support through grant AST121159. GB acknowledges partial support from PRIN-INAF 2014. DR was supported by the National Research Foundation of Korea through grants 2016R1A5A1013277 and 2014M1A7A1A03029872.

\bibliographystyle{mnras}


\appendix

\section*{A1. Tests of algorithms}
\subsection*{A1.1 Spatial filtering for turbulence}
\label{subsec:A1}
Similar to \citet{va12filter}, we ran several tests of our iterative filtering procedure for turbulence (Sec.\ref{subsec:turbo}) over 
control turbulent boxes with predefined outer scales and slopes for the power spectrum. 
In the example given here, we analysed a purely solenoidal velocity field on a 
$100^3$ grid, first  generating a vector potential and then computing its curl. The vector potential was drawn from $3 \leq k \leq 50$ and had a power-law spectrum with a slope $E_v(k) \propto k^{-5/3}$. 
We then ran  our algorithm with the same set of parameters as the analysis in the paper and checked whether the input velocity field was correctly recovered. 
The panels in Figure \ref{fig:figA2} present the performance of this test, comparing the power spectrum and the 2$^{nd}$ order structure function of both input and reconstructed velocity field using our filtering procedure \ref{subsec:turbo}. 

\begin{figure}
\includegraphics[width=0.45\textwidth]{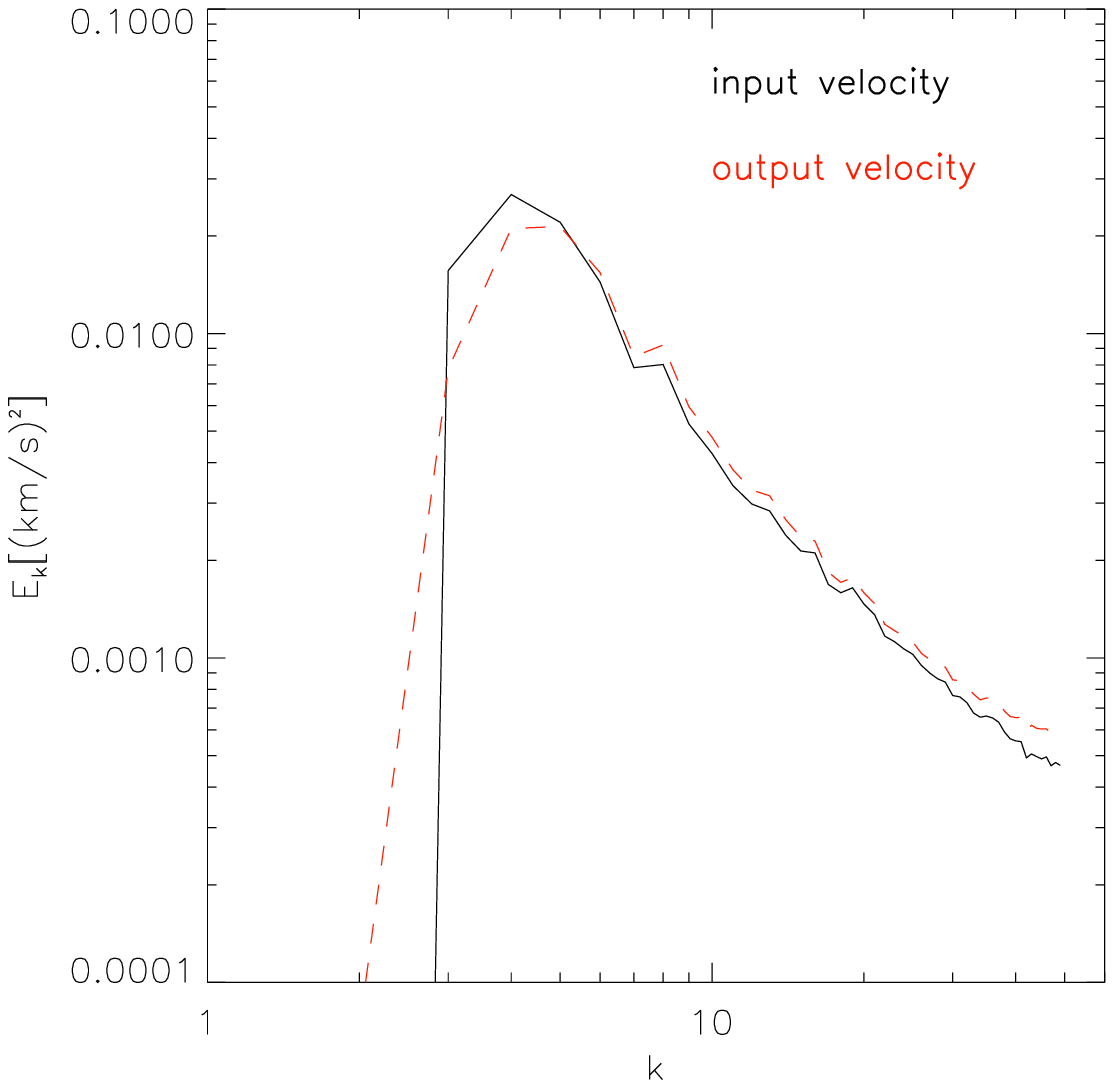}
\includegraphics[width=0.45 \textwidth]{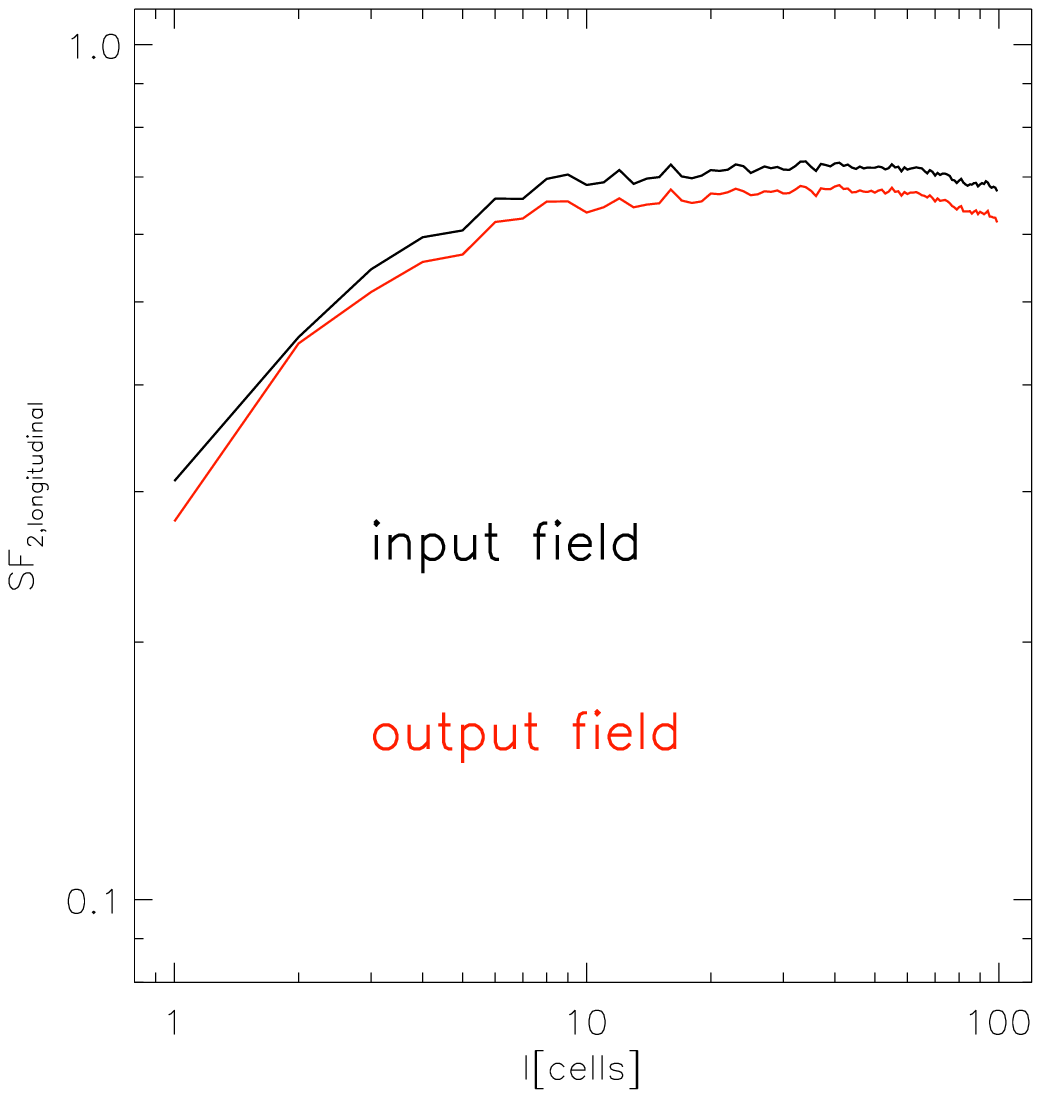}
\caption{Top: velocity field power spectra for the input velocity of our test (black) and for the output velocity field as reconstructed by our iterative filter (red). Bottom: 2nd order longitudinal structure functions for the same test field, with identical meaning of colors.}
\label{fig:figA2}
\end{figure}


The results are representative of the performances of the filter in this kind of test, which also applies by construction when regular fields on large scales are imposed on the 
setup \citep[see, e.g.][for similar tests]{va12filter}). 
Both the outer injection scale of the flow and the spectral behaviour of the fluctuating field are accurately recovered by our filtering procedure.
The slope of the power spectrum and of the structure function are recovered within $\sim 5-10\%$  accuracy.
In these tests, typically only $\sim 10\%$ percent of the kinetic energy in the fluctuating field is not recovered after the filtering procedure. That is due to a misidentification with a large-scale smooth flow, which can occur when the outer correlation scales of the flow are of the order or larger than the computing domain. (Hence, they mimic a large scale regular component.)

On the other hand, the kinetic energy flux within the cascade is captured extremely well.  For the main goals of our ICM analysis this small difference plays only a minor role, and  for the purpose of our analysis in the main paper this filtering technique is suitable for capturing the most important small-scale turbulent features of the flow. 

\subsection*{A1.2 Hodge-Helmoltz modes decomposition}
\label{subsec:A2}

We tested the ability of the Helmholtz-Hodge (``HH'') procedures employed in Sec.\ref{subsec:helm} to decompose blended solenoidal/compressive velocity fields correctly and in comparison to a simple, ``straw man'' alternative. The basis for the HH procedure is that a vector field with suitable asymptotic/boundary behaviors can be expressed in a form $\vec v = -\nabla \phi + \nabla\times \vec a = \vec v_c + \vec v_s$ \citep[e.g.,][]{arfken95}. In particular, assuming the vector field, $\vec v$, has a Fourier transform, $\vec V_k$, 
one can obtain the two components, $\vec v_c$ and $\vec v_s$. The algorithm applied in \ref{subsec:helm}  (and which we call here Method 1) assumed explicitly that $\vec V_k =  i \vec k \Phi + i \vec k \times \vec A = \vec V_{k,c} + \vec V_{k,s}$ (where capital letters in this discussion generally refer to Fourier transforms of lower case spatial fields); that is, that the Fourier transform, $\vec V_{k,c} = \vec k (\vec k \cdot \vec V_k)/|k|^2$, can be found by projection of $V_k$ onto $\vec k$ and that $\vec V_{k,s} = -\vec k\times(\vec k\times\vec V_k)/|k|^2$ can be found normal to $\vec k$. The alternate approach (which we call Method 2 here), instead applied fourth order finite difference methods to estimate $d = -\nabla^2\phi = \nabla\cdot\vec v$, then obtained $\Phi = D/k^2$. The inverse Fourier transform yielded $\phi$, while $\vec v_c = -\nabla \phi$ was obtained  from $\phi$ using fourth order finite differences.  We note that use of high order differences for spatial derivatives is especially important in attempts to extract compressive component variations close to the Nyquist frequency, since those difference algorithms can construct exact derivatives of polynomials up to the order of the differencing. Equivalent procedures to those employed to find $\vec v_c$ could be used to find $\vec v_s$ directly, of course. In these tests, however, we found $\vec v_c$ explicitly, then obtained $\vec v_s$ as $\vec v_s = \vec v  - \vec v_c$.

\begin{figure*}
\includegraphics[width=0.85\textwidth]{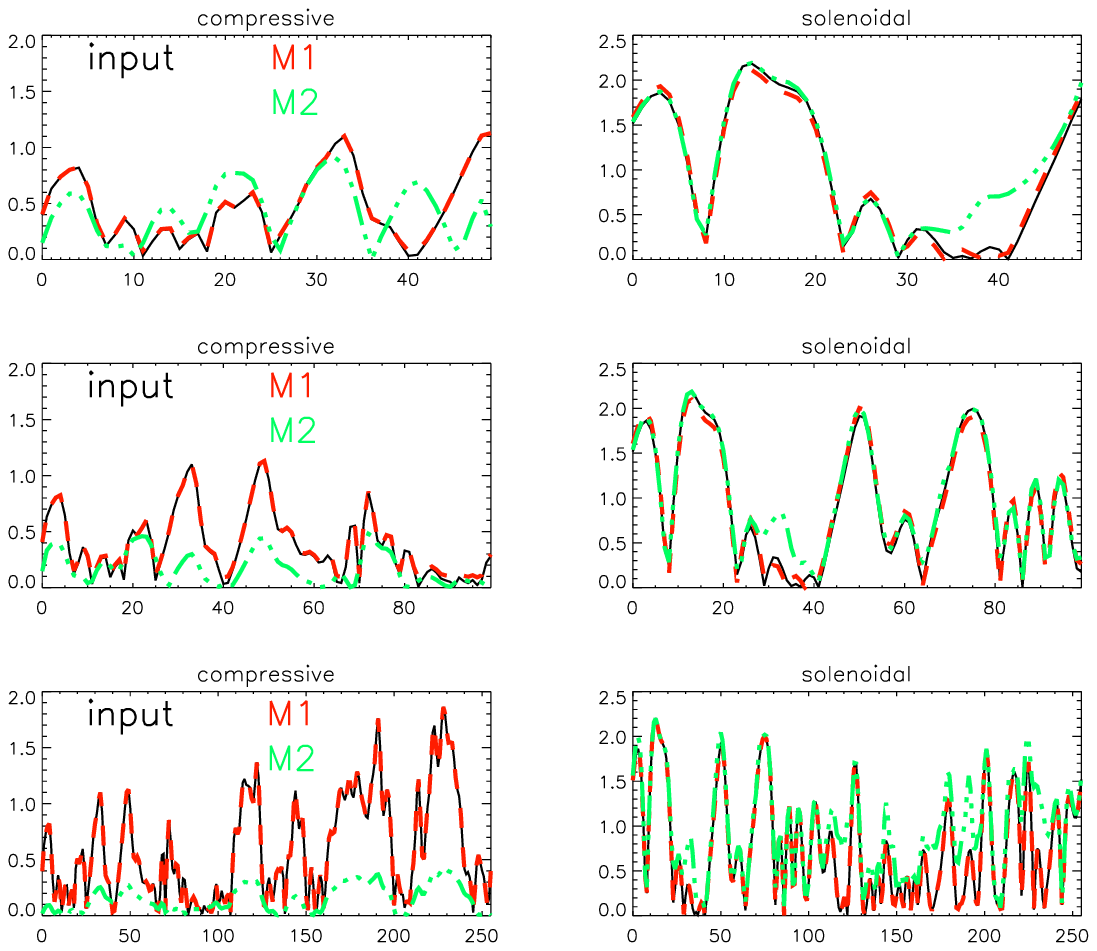}
\caption{Helmholtz decomposition into compressive (left) and
solenoidal (right) component magnitudes of a 3D velocity  field plotted
along x using Method 1(red) and Method 2 (green), as explained in the text. The constructed, "input" components are shown in black. The different rows correspond
to different sized portions of the box: full $256^3$ zone, periodic box; middle: $100^3$ zone sub-volume; top: $50^3$ zone sub-volume. }
\label{fig:figA4}
\end{figure*}

To carry out the HH tests we first constructed idealized 3D velocity fields in a cubic box of arbitrary length, $L$, with both compressive, $\vec{v_c}$, and solenoidal,$\vec{v_s}$, velocity components using discrete Fourier sums; $\vec v(\vec x) = \sum \vec V_k cos(\vec k \cdot \vec x + \psi_k)$, with $0\le \psi_k\le 2\pi$ selected from random deviates. 
For simplicity we aligned all the component wave vectors, $\vec k$ along the $x$ axis; that is $\vec{k} = k_x \hat{x}$, with $|k| = 2\pi n/L$ and $|n|>0$. Then, of course, the velocity fields were periodic on length, $L$, while each compressive (solenoidal) Fourier component satisfied $\vec k \times \vec V_k = 0$ ( $\vec k \cdot\vec V_k = 0$). The compressive and solenoidal fields were given distinct, non-vanishing power spectra over distinct ranges in $k_x$, $\vec v_c$ and $\vec v_s$ can be obtained from $\Phi$ and $\vec A$, respectively, as outlined in the previous paragraph. Note that there is no power in $n = 0$ modes. 

The constructed velocity field was evaluated on $256^3$ uniform spatial grid points spanning $L$. Although the constructed vector fields were periodic over the full, $256^3$ domain, velocity fields in our cluster simulations generally are not periodic in domains of interest. Therefore, we tested the accuracy of both HH algorithms to recover correct velocity information from nonperiodic, $100^3$ and $50^3$ cell subvolumes.

An example test velocity field is illustrated by the solid, black lines in Fig.\ref{fig:figA4}. The compressive (solenoidal) component is on the left (right). The bottom panels show the velocity distributions across the full domain, while the middle (top) row show them in a $100^3$ ($50^3$)  subvolume. The case shown in Fig.\ref{fig:figA4} is for a compressive component with modes spanning $2\leq k_x \leq 128$ with a $E(k) \propto k^{-2}$ spectrum, and a solenoidal component with a $\propto k^{-5/3}$ spectrum in modes spanning $4 \leq k_x \leq 32$. 
 
Fig. \ref{fig:figA4} shows the outcomes of the Method 1 and Method 2 HH decompositions of this velocity field with  Method 1 shown in red and Method 2 in green. Both on the full, periodic volume and the non-periodic subvolumes the Method 1 reconstructions are almost identical to the input velocity field. That is, this method proved to be quite accurate. In contrast Method 2 clearly misses substantial power in the compressional component, despite the use of high order difference algorithms to estimate spatial derivatives. In conclusion, then, the Method 1 HH decomposition that depends entirely on FFTs is a reliable approach to obtain the compressive and solenoidal velocity components, even when the domain of interest is not periodic. In contrast, the use of finite differences to avoid FFT extraction of non-periodic spatial gradient information was not successful. 

\end{document}